\documentclass[usenatbib]{mn2e}
\input{psfig}
\usepackage{natbib}  
\usepackage{varioref}

\begin{document} 
\def\lsim{\mathrel{\hbox{\rlap{\hbox{\lower4pt\hbox{$\sim$}}}\hbox{$<$}}}}
\def\gsim{\mathrel{\hbox{\rlap{\hbox{\lower4pt\hbox{$\sim$}}}\hbox{$>$}}}}
\def\simlt{\mathrel{\rlap{\lower 3pt\hbox{$\sim$}}
        \raise 2.0pt\hbox{$<$}}}
\def\simgt{\mathrel{\rlap{\lower 3pt\hbox{$\sim$}}
        \raise 2.0pt\hbox{$>$}}}

\title[Spectral and morphological properties of quasar hosts in SPH simulations of AGN feeding by mergers]
{Spectral and morphological properties of quasar hosts in SPH simulations of AGN feeding by mergers}
\author[A. Cattaneo, F. Combes, S. Colombi, E. Bertin, A.-L. Melchior]
{A.~Cattaneo $^{1,2,3}$, F. Combes $^{4,5}$, S. Colombi $^{3,6}$, E. Bertin $^3$, A.-L. Melchior $^{4,5,6}$\\
$^1$Institut d'Astrophysique de Paris, 98bis Boulevard Arago, 75014 Paris, France\\
$^2$Racah Institute of Physics, The Hebrew University, 91904 Jerusalem, Israel\\
$^3$Astrophysikalisches Institut Potsdam, an der Sternwarte 16, 14482, Potsdam, Germany\\
$^4$Observatoire de Paris, 61 Avenue de l'Observatoire, 75014 Paris, France\\
$^5$Galaxy Formation (GalFor), HORIZON, CNRS, France\\
$^6$Numerical Investigations in Cosmology (N.I.C.), HORIZON, CNRS, France}

\maketitle 
\begin{abstract}
We present a method for generating virtual observations from smoothed-particle-hydrodynamics (SPH) 
simulations.
This method includes stellar population synthesis models and the reprocessing of starlight by dust
to produce realistic galaxy images.
We apply this method and simulate the merging of two identical giant Sa galaxies
($M_{\rm disc}=10^{11}M_\odot$, $M_{\rm spheroid}=2.5\times 10^{10}M_\odot$).
The merger remnant is an elliptical galaxy ($M_{\rm spheroid}\simeq 1.3\times 10^{11}M_\odot$,
$M_{\rm disc}\simeq 7.4\times 10^{10}M_\odot$).
The merger concentrates the gas content of the two galaxies into the nuclear region.
The gas that flows into the nuclear region refuels the central black holes of the merging galaxies.
We follow the refuelling of the black holes during the merger semi-analytically. 

In the simulation presented in this article, the black holes grow from $3\times 10^7\,M_\odot$
to $1.8\times 10^8M_\odot$, with a peak AGN luminosity of $M_B\sim -23.7$. 
We study how the morphological and spectral properties of the system evolve during the merger and
work out the predictions of this scenario for the properties of host galaxies during the active phase.
The peak of AGN activity coincides with the merging of the two galactic nuclei and 
occurs at a stage when the remnant looks like a lenticular galaxy.
The simulation predicts the formation of a circumnuclear starburst ring/dusty torus with an opening angle of 
$30^\circ-40^\circ$ and made of clouds with $n_{\rm H}=10^{24}{\rm cm}^{-2}$. 
The average optical depth of the torus is quite high, but the obscuring medium is patchy, so that there
still exist lines of sight where the AGN is visible in a nearly edge-on view.
For the same reason, there are lines of sight where the AGN is completely obscured in the face-on view.
\end{abstract}

\begin{keywords}
galaxies: formation, interactions, active, nuclei -- quasars: general -- methods: N-body simulations
\end{keywords}

\section{Introduction}

\citet{toomre_toomre72} suggested that mergers can convert spiral galaxies into elliptical ones. 
They also remarked that galaxy interactions can determine strong bar instabilities, 
trigger starbursts and feed gas to active galactic nuclei (AGN). 
This paradigm, which links the formation of elliptical galaxies and AGN to mergers, 
has derived its strength from its success in smoothed-particle-hydrodynamics (SPH) simulations of 
galaxy mergers (i.e. \citealp{barnes_hernquist91,mihos_hernquist94,barnes_hernquist96,mihos_hernquist96})
and in semi-analytic models of galaxy formation 
(i.e. \citealp{kauffmann_etal93,kauffmann_charlot98,kauffmann_haehnelt00,cattaneo01}).
The discovery of ultra-luminous infrared galaxies (ULIRGs) with the infra-red astronomical satellite IRAS
provided further support for the merger scenario. 
These galaxies release most of their power in the IR because most of their light is reprocessed by dust. 
A significant fraction of these objects host AGN and many are observed to be part of interacting
or merging systems (i.e. \citealp{bushouse_etal02}).

Nevertheless, from an observational point of view, the link of AGN with galaxy interactions is highly 
controversial. Since the earliest detections of quasar host galaxies with the Canada-France-Hawaii Telescope,
there has been a claim that $>30\%$ quasar hosts are currently interacting with a companion 
\citep{hutchings_campbell83}. 
\citet{xilouris_papadakis02} fitted de Vaucouleurs profiles to HST images of faint AGN and normal galaxies from the local Universe.
After subtracting
their fits from the data, they concluded that all active galaxies show significant structure in their inner 100\,pc and 
1\,kpc regions, contrary to quiescent early-type galaxies, which show no structure at all.
However, other groups find that the isophotal profiles of quasar hosts are 
consistent with the de Vaucouleurs law and that their colours are those of old, passively evolving stellar 
populations, while interactions are no more common than in normal galaxies with similar characteristics 
\citep{dunlop_etal03}.

There is now evidence that AGN seen through their intense optical/UV continuum (big blue bump) and broad emission line
spectra (Type I quasars) are just the tip of the iceberg of the AGN phenomenon 
\citep{brandt_etal01,barger_etal02,ueda_etal03,sazonov_etal04}) at least in terms of the number of objects.
\citet{yu_tremaine02} used \citet{soltan82} and \citet{chokshi_turner92}'s
method to infer the cosmic density of supermassive black holes due to optical 
quasars from the quasar luminosity function.
They inferred a value of $2.1\times 10^5M_\odot{\rm Mpc}^{-3}$, which they compared with their best estimate of the local
density of supermassive black holes, $(2.5\pm 0.4)\times 10^5M_\odot{\rm Mpc}^{-3}$, concluding that there is limited need for
obscured or radiatively inefficient accretion.
However, Barger et al. (2005) have shown that extrapolating a fit to the quasar luminosity function outside the measured magnitude
range can overestimate the  cosmic density due to optical quasars by a factor of 1.75.
Moreover, \citet{marconi_etal04} disagree with the local density of supermassive black holes estimated by \citet{yu_tremaine02}
and propose a value almost twice as large. 
Unification by orientation \citep{antonucci_miller85} 
has proven a successful paradigm to explain the type I/II distinction.
In this paradigm, AGN are surrounded by dusty tori.
From a pole-on view, we see the naked AGN and the object is classified as type I.
From an equatorial view, the dusty torus obscures the central engine and all we can see is reverberated light/narrow
line emission from clouds illuminated by the AGN. In this case the object is classified as type II.

Studying the host galaxies of type II AGN is easier because there is not the problem of subtracting 
the light of the AGN. Nature does it for us.
\citet{kauffmann_etal03} analysed the spectra of 22 623 narrow line AGN from the Sloan Digital Sky Survey. 
By using the discontinuity at $0.4\,\mu{\rm m}$ as an indicator of recent star formation and 
the strength of the O III $0.5007\,\mu{\rm m}$ emission line as an indicator of AGN power, 
they concluded that low luminosity AGN are hosted by normal early-type galaxies.
The hosts of bright AGN are galaxies with the high surface densities of early-type galaxies, but bluer
colours indicating recent star formation. They claim that this star formation is not concentrated to the
nuclear region. This finding appears to disagree with that of \citet{dunlop_etal03} and questions 
the paradigm of unification by orientation. I.e., are the hosts of broad and narrow line AGN drawn from 
the same population?
If starbursts on the scale of the host galaxy contribute to obscure the central engine and the broad line 
region, then type I samples may be biased toward AGN in late-stage or gas-poor mergers.

Here we use numerical simulations i) to work out the expected properties of quasar hosts in the merging 
scenario and ii) to understand the difficulties and selection effects that are present when we try to 
derive properties of quasar hosts from observations. 
For this purpose, we run hydrodynamic N-body simulations of galaxy mergers, 
where we incorporate the effects of absorption by dust and a semi-analytic treatment of the growth of 
supermassive black holes, interfaced with visualising tools, which convert the outputs of the simulations
into mock data (including instrumental diffraction, sky noise, etc.)

The hydrodynamic simulations use the three-component (stars, gas and dark matter) SPH tree-code 
GalMer, \citep{combes_melchior02} described in Section 2. 
The modelling of the stellar spectral energy distribution (SED) and of the reprocessing of star light by 
dust is performed after the simulations, through an interface that manipulates the simulations' outputs (Section 3).
The same is true for the growth of the central black hole, which is computed semi-analytically from the rate at
which gas falls into the central $100\,$pc.

Section 5 presents the results of our analysis of the outputs of the simulation.
In this analysis, we concentrate on three questions:
i) the relation between the global star formation history of the host galaxy and star formation/black hole accretion in the
galactic nucleus,
ii) the relation  between the spectro-morphological evolution of the host galaxy and the growth of the central black hole,
and iii) the geometry, the optical depth and the covering factors of the absorbing materials.
In Section 6, we summarise the main new results of this work.

This paper is part of a project to build a library of simulated galaxy mergers.
Here, we only consider one simulation where we merge two identical Sa galaxies.
In a future publication, we shall discuss how the results of this article depend on the merging parameters
and on the properties of the merging galaxies.


\section{The hydrodynamic simulations}

\subsection{The GalMer SPH code}
 
Gravitational hydrodynamics is simulated with the GalMer SPH tree-code, which is a development of the 
code used by \citet{combes_melchior02}. 
In GalMer, galaxies are composed of three kinds of particles: stellar particles, dark matter particles
and hybrid particles. 
At the beginning of a simulation, a hybrid particle is made entirely of gas. 
We assume that the gas is isothermal and cools very efficiently.
We calculate pressure forces for an isothermal gas with a temperature of 10,000\,K. This is approximately the temperature
where there is strong change in the cooling curves \citep{blumenthal_etal84}.
The adopted star formation rate is given by the modified Schmidt law:
$$\dot{\rho}_*= 8.33\times 10^7\,M_\odot{\rm\,kpc^{-3}Myr^{-1}}\times$$
\begin{equation}
\times\left({\rho_{\rm g}\over 10^9\,M_\odot{\rm\,kpc^{-3}}}\right)^{1.2}
\left({\sigma\over 100{\rm\,km\,s}^{-1}}\right)^\eta,
\end{equation}
where $\rho_{\rm g}$ is the density of the gas and 
$\sigma$ is the local velocity dispersion of the hybrid particles. 
We take $\eta=1.5$ for regions where the divergence is negative and the gas is shocked, while we 
take $\eta=0$ for regions where the gas is expanding. 
When the gas content of a hybrid particle drops below 5$\%$, the hybrid particle 
is turned into a stellar particle and its gas content is spread among the neighbouring
hybrid particles. 
It is useful to think of a hybrid particle as a model for a giant molecular cloud, 
which evolves into a star cluster once all the gas has been locked into stars.

\subsection{The initial conditions for the merging galaxies}

In this first paper, we concentrate on the results of one simulation, in which we merge two identical 
galaxies. Each of the two galaxies is made of four components: the dark matter halo (dark matter particles),
the stellar disc (stellar particles), the gas disc (hybrid particles) and the bulge (stellar particles).
Each of the four components is a random realisation of a Miyamoto-Nagai density profile 
\citep{miyamoto_nagai75}, which, in cylindrical coordinates $r,\phi,z$, takes the form:
\begin{equation}
\rho(r,z)={b^2M\over 4\pi}
{ar^2+[a+3(z^2+b^2)^{1/2}][a+(z^2+b^2)^{1/2}]^2 \over
\{r^2+[a+(z^2+b^2)^{1/2}]^2\}^{5/2}(z^2+b^2)^{3/2}}.
\end{equation}
$M$ is the total mass of the component, while $a$ and $b$ are two characteristic length-scales.
The Miyamoto-Nagai density distribution tends to a Plummer sphere with radius $b$ in
the limit $a\rightarrow 0$ and to a Kuzmin disc with radius $a$ in the limit $b\rightarrow 0$.
These density distributions are discussed at length in the textbook by \citet{binney_tremaine87}.
Table 1 gives the mass $M$, the number of particles $N_{\rm p}$ and the characteristic scale-lengths 
$a$ and $b$ for the each of the four components. 
The masses and the numbers of particles that are given in Table 1 are the values for one galaxy.
\begin{table}
\caption{Initial galaxy parameters in the SPH simulation}
\begin{tabular}{l|l|c|c|c}
Component    &  $M$ ($M_\odot$)  & $N_{\rm p}$  & a\,(kpc) & b\,(kpc) \\
\hline
DM halo      & $1.25\cdot 10^{11}$& 40,000     &    0     &  50     \\
Stellar disc & $1.00\cdot 10^{11}$& 48,000     &    4     &   0.5   \\
Gas disc     & $1.00\cdot 10^{10}$& 20,000     &    5     &   0.2   \\
Bulge        & $2.50\cdot 10^{10}$& 12,000     &    0     &   2     \\
\hline
\end{tabular}
\end{table}
The total number of particles is 240,000.

\subsection{The SPH outputs}

We save the outputs of the SPH simulations at intervals of 25\,Myr. The five snapshots in Figs.~1-2 
convey an idea of the merging dynamics. Soon after the simulation has started, the discs become
unstable in the bar mode. The bars in the gas discs are stronger than the bars in the stellar discs 
and are accompanied by more violent inflows along the bar direction.

The galaxy labelled as galaxy 1 is the one that is above in the
100\,Myr snapshot and below in the 200\,Myr snapshot (Fig.~1). 
The two galaxies are identical, but
it is to this one that we refer
when we give information pertaining to one galaxy only (zoom, central star formation, etc.).

\section{From SPH simulations to mock astronomical data}
\begin{figure*}
\noindent
\begin{minipage}{8.6cm}
  \centerline{\hbox{
      \psfig{figure=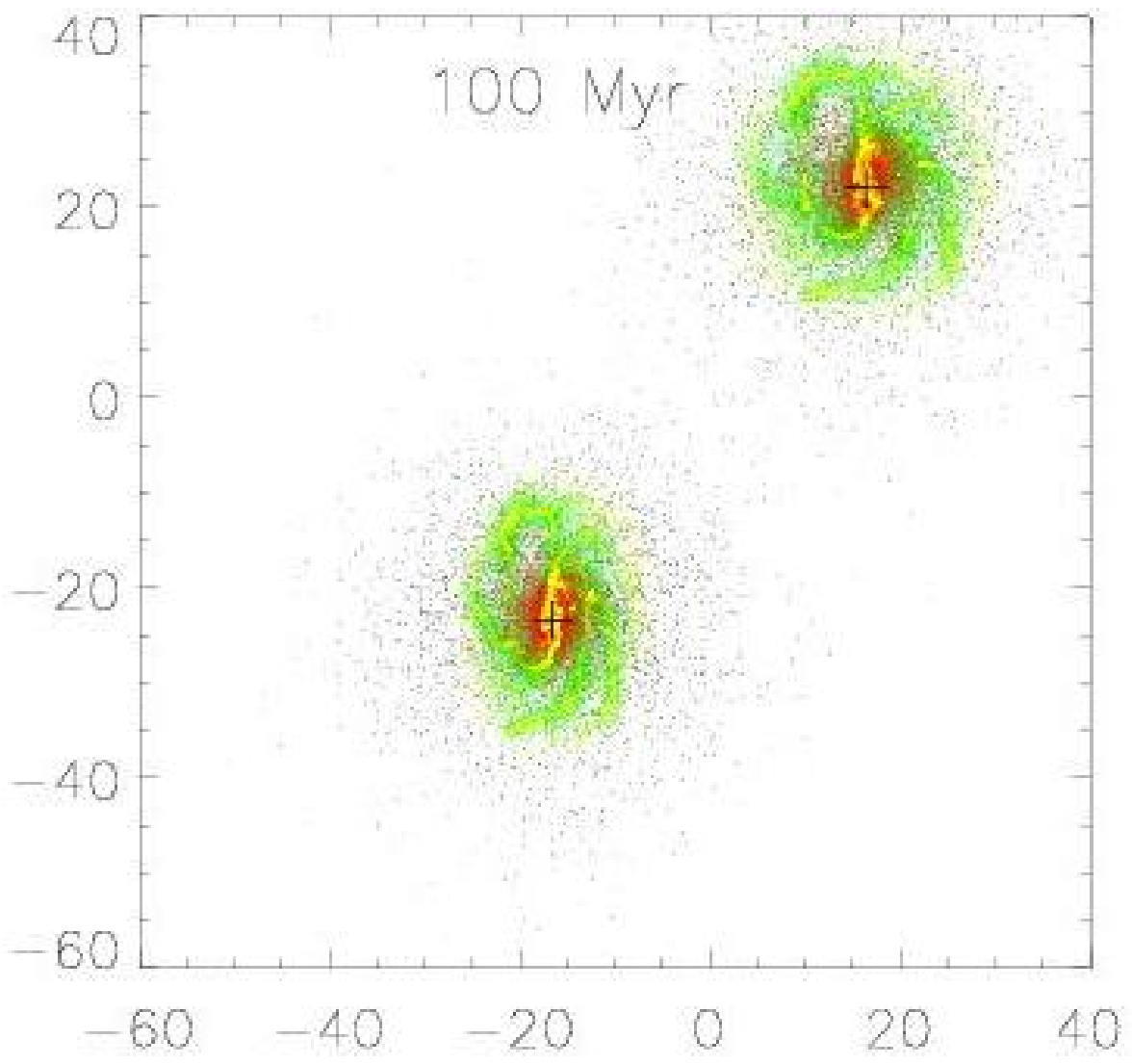,height=7.5cm,angle=0}
  }}
\end{minipage}\    \
\begin{minipage}{8.6cm}
  \centerline{\hbox{
      \psfig{figure=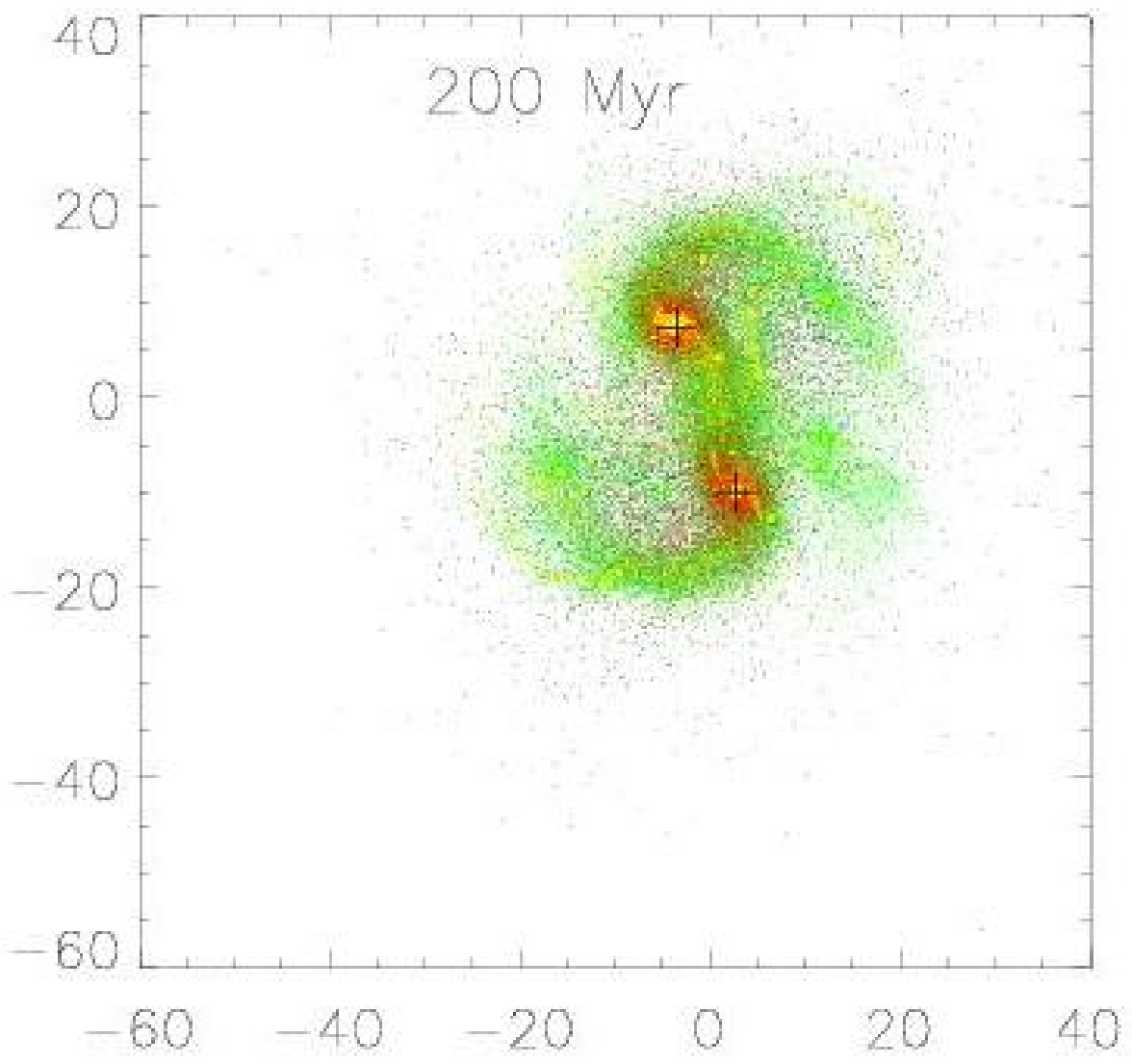,height=7.5cm,angle=0}
  }}
\end{minipage}\    \
\begin{minipage}{8.6cm}
  \centerline{\hbox{
      \psfig{figure=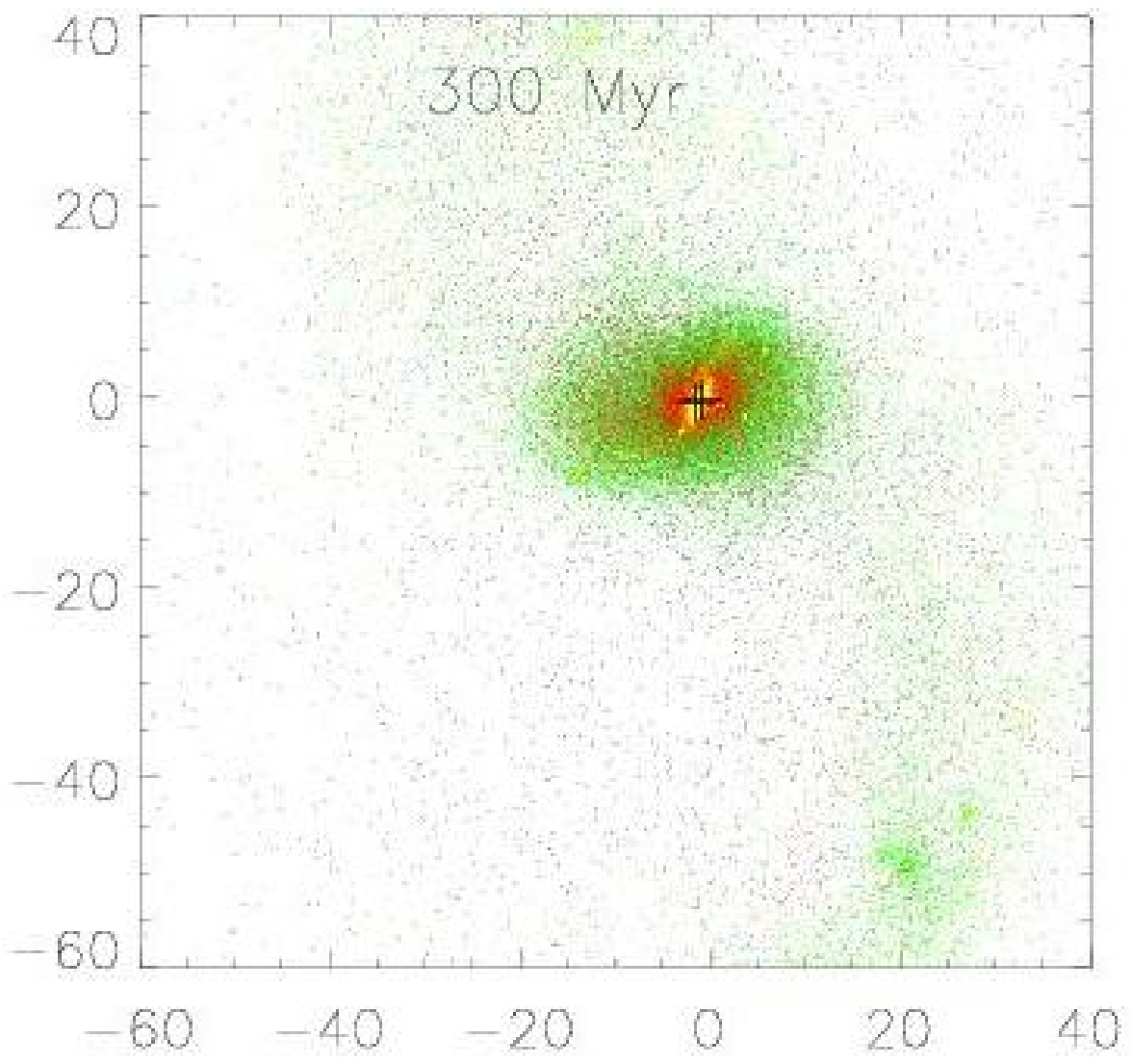,height=7.5cm,angle=0}
   }}
\end{minipage}\    \
\begin{minipage}{8.6cm}
  \centerline{\hbox{
      \psfig{figure=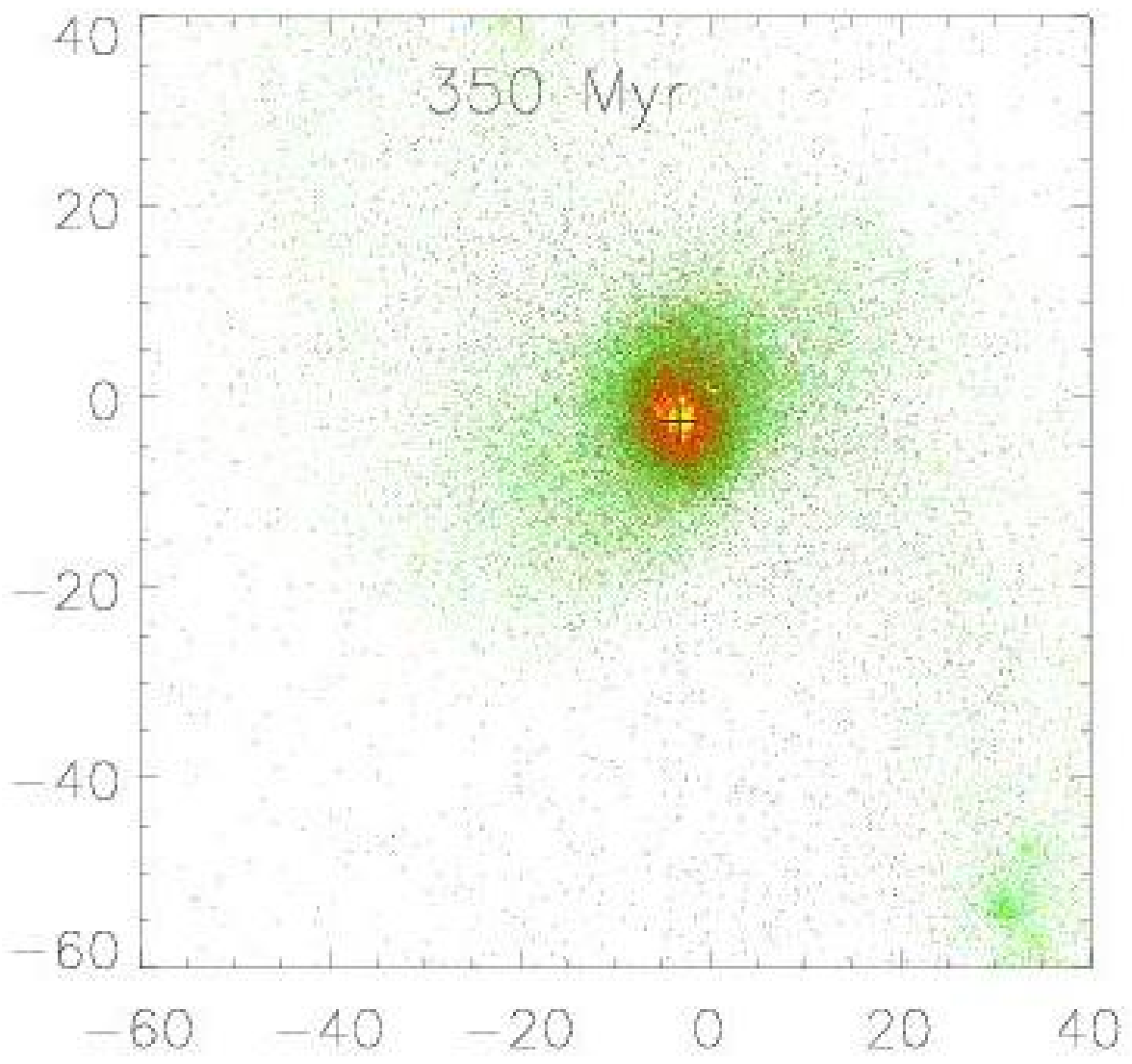,height=7.5cm,angle=0}
  }}
\end{minipage}\    \
\begin{minipage}{8.6cm}
  \centerline{\hbox{
      \psfig{figure=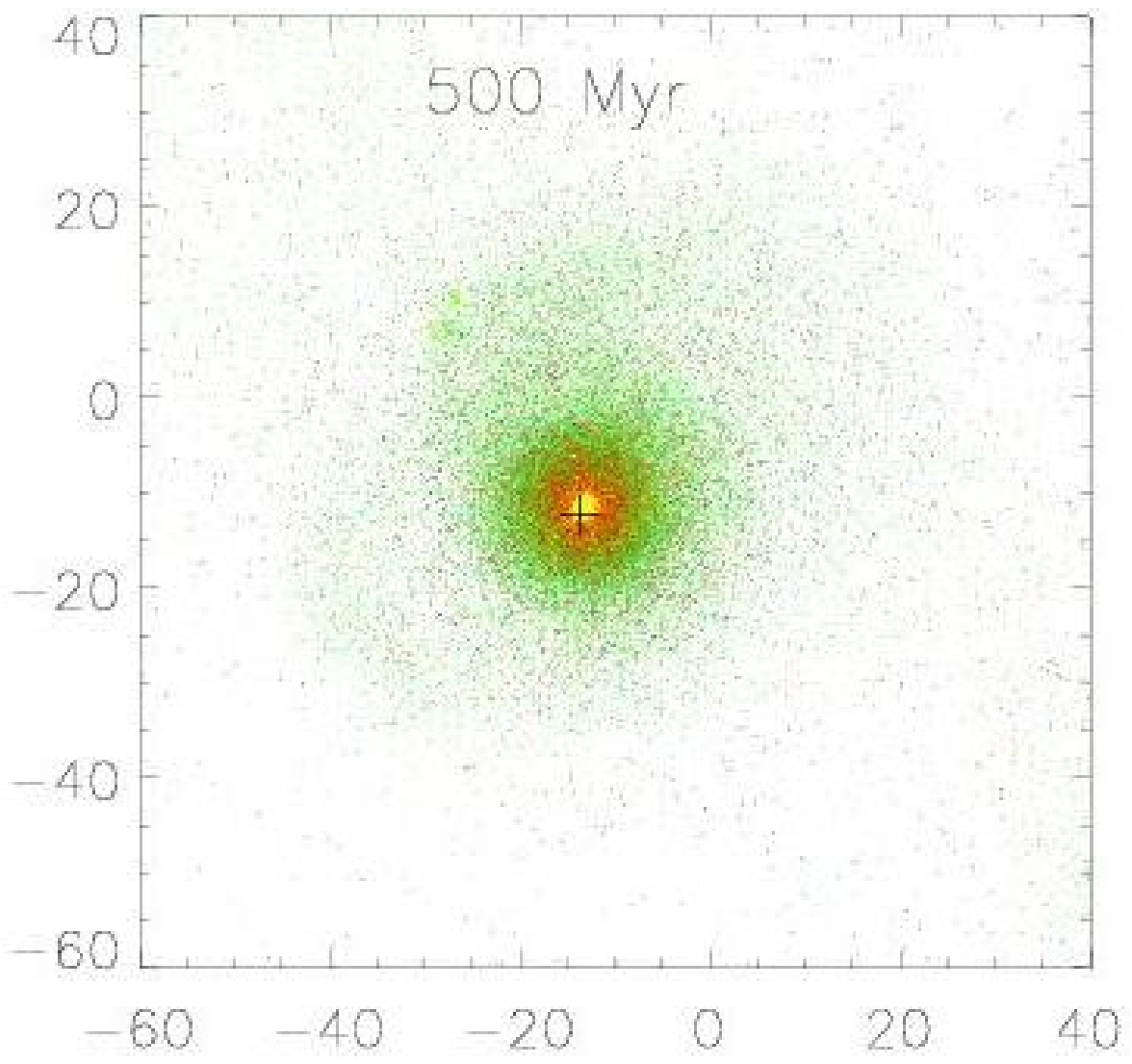,height=7.5cm,angle=0}
  }}
\end{minipage}\    \
\begin{minipage}{8.6cm}
  \centerline{\hbox{
      \psfig{figure=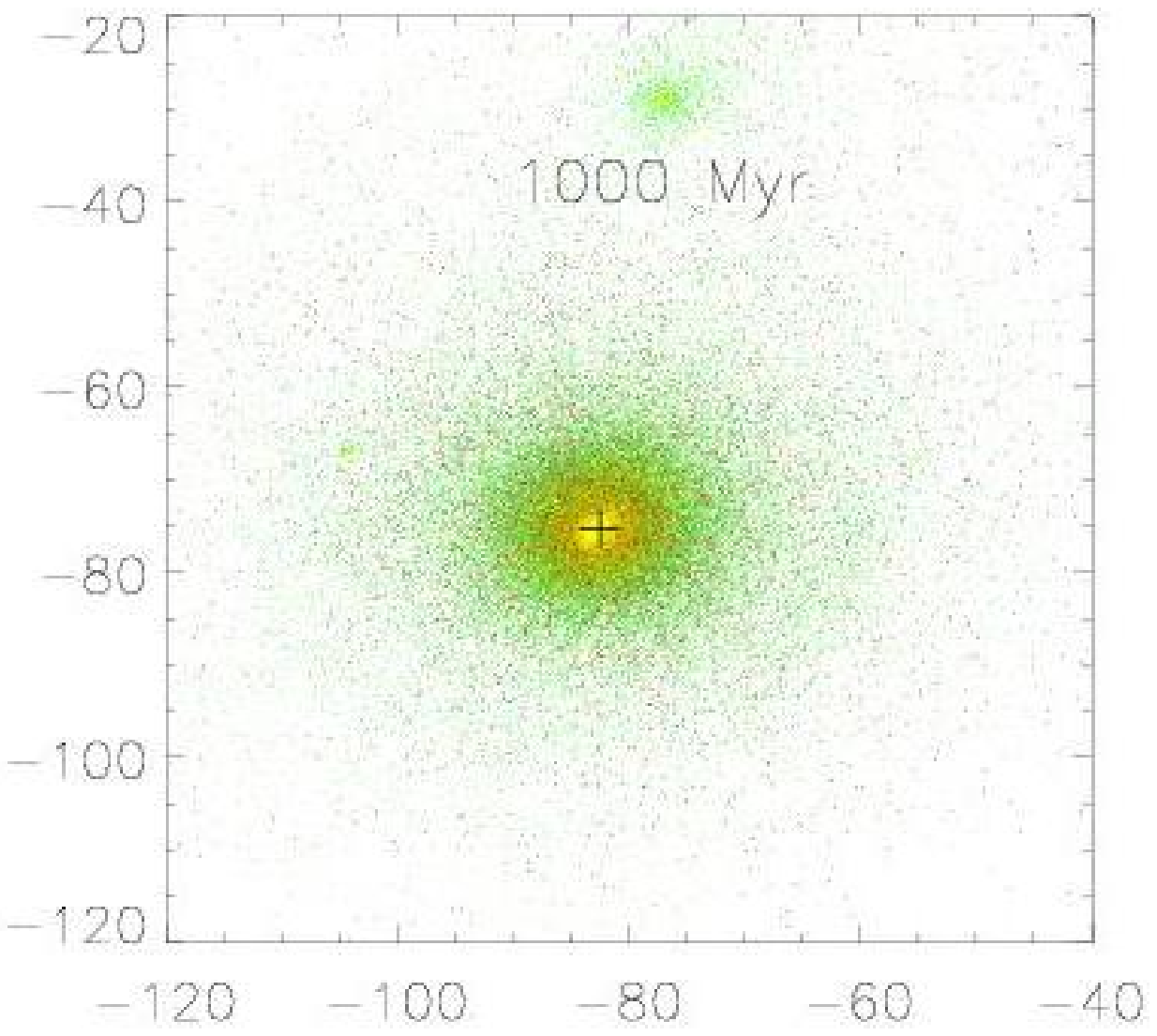,height=7.5cm,angle=0}
  }}
\end{minipage}
\caption{A sequence of five snapshots from the GalMer SPH simulation. 
The time elapsed from the beginning of the simulation is printed on each snapshot.
The units on the axis are kpc. Black is dark matter, green is old disc stars,
red is bulge stars and yellow is gas/young stars.} 
\end{figure*}

\begin{figure*}
\noindent
\begin{minipage}{8.6cm}
  \centerline{\hbox{
      \psfig{figure=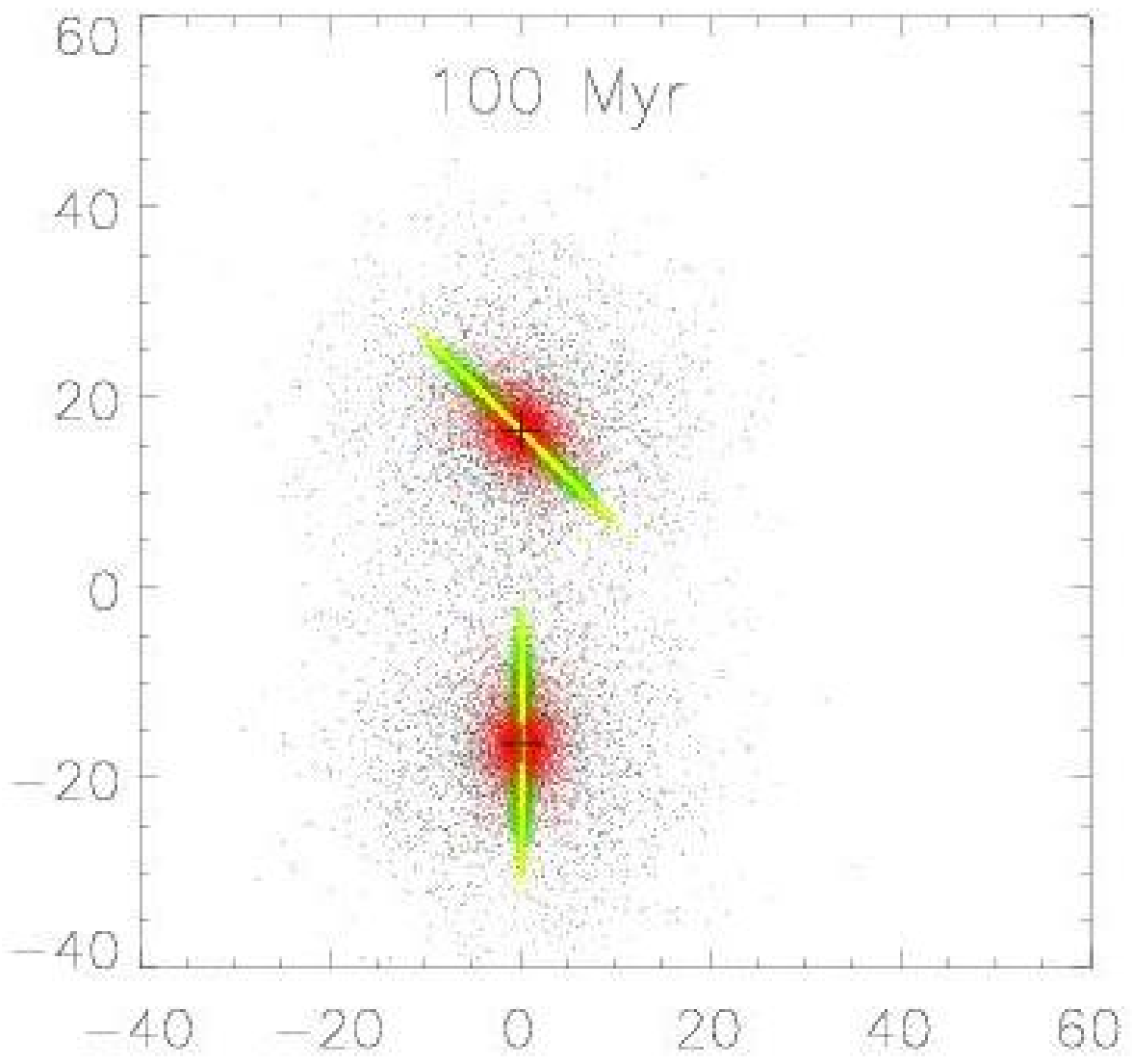,height=7.5cm,angle=0}
  }}
\end{minipage}\    \
%
%
\begin{minipage}{8.6cm}
  \centerline{\hbox{
      \psfig{figure=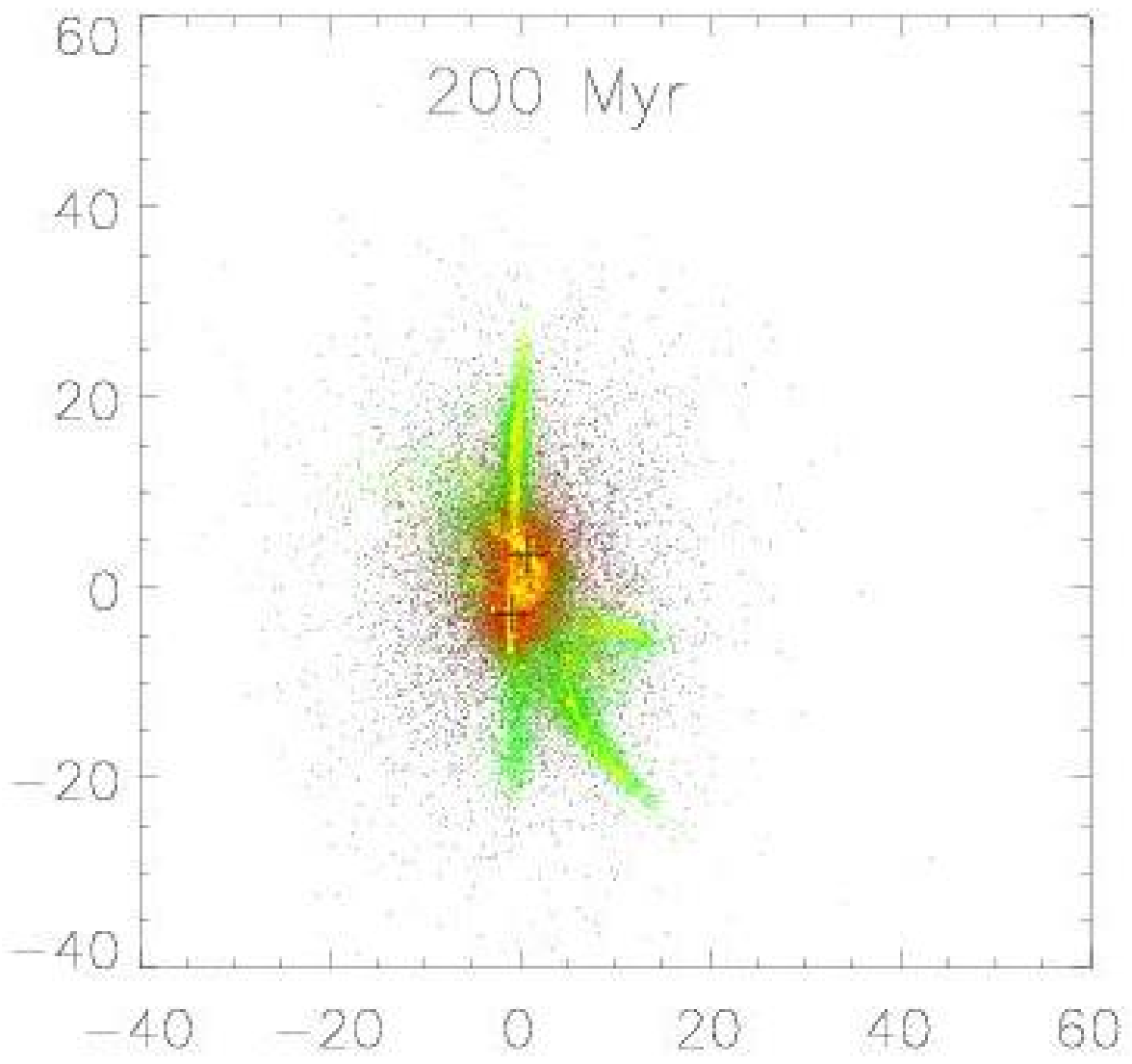,height=7.5cm,angle=0}
  }}
\end{minipage}\    \
\begin{minipage}{8.6cm}
  \centerline{\hbox{
      \psfig{figure=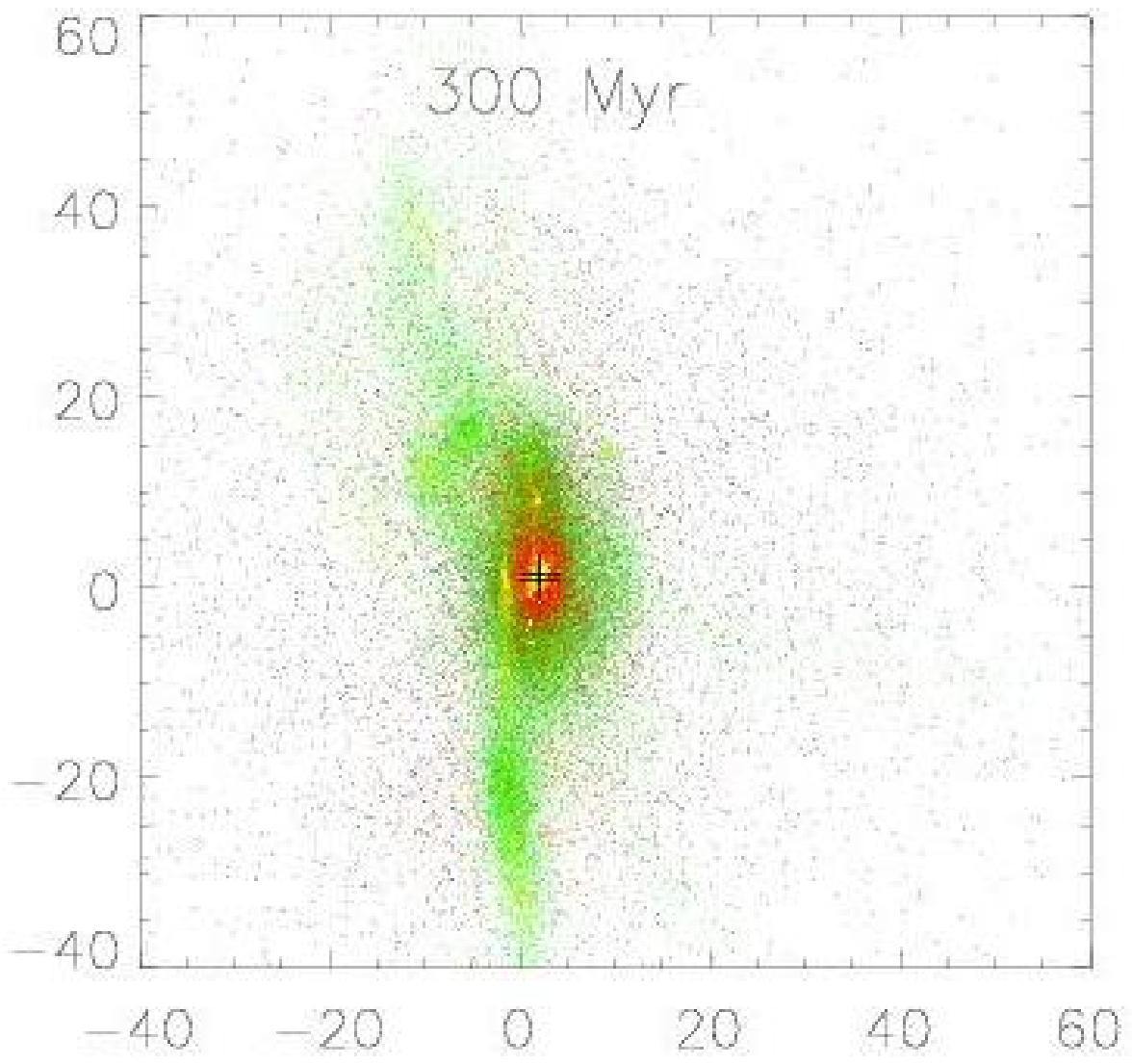,height=7.5cm,angle=0}
  }}
\end{minipage}\    \
\begin{minipage}{8.6cm}
  \centerline{\hbox{
      \psfig{figure=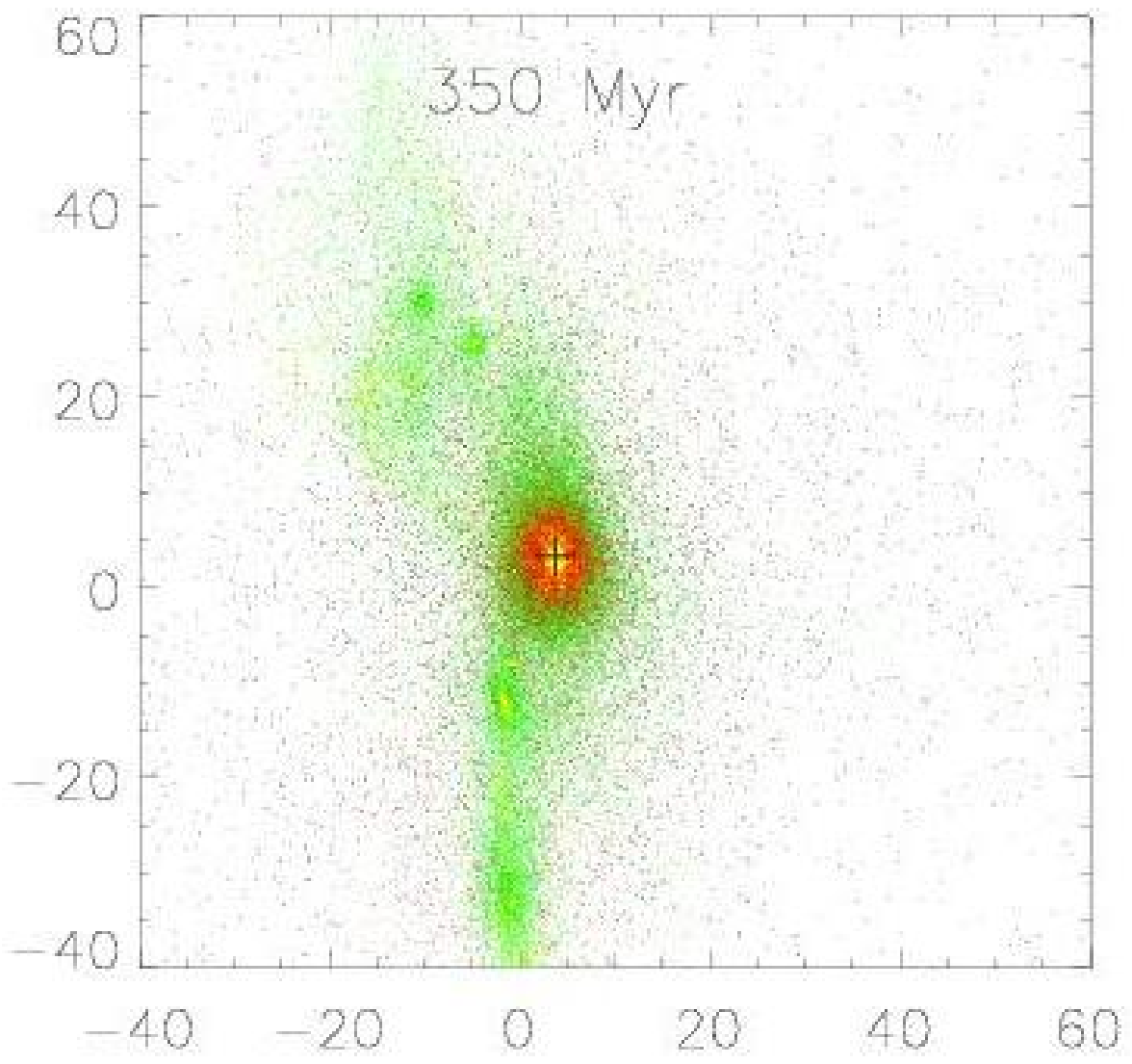,height=7.5cm,angle=0}
  }}
\end{minipage}\    \
\begin{minipage}{8.6cm}
  \centerline{\hbox{
      \psfig{figure=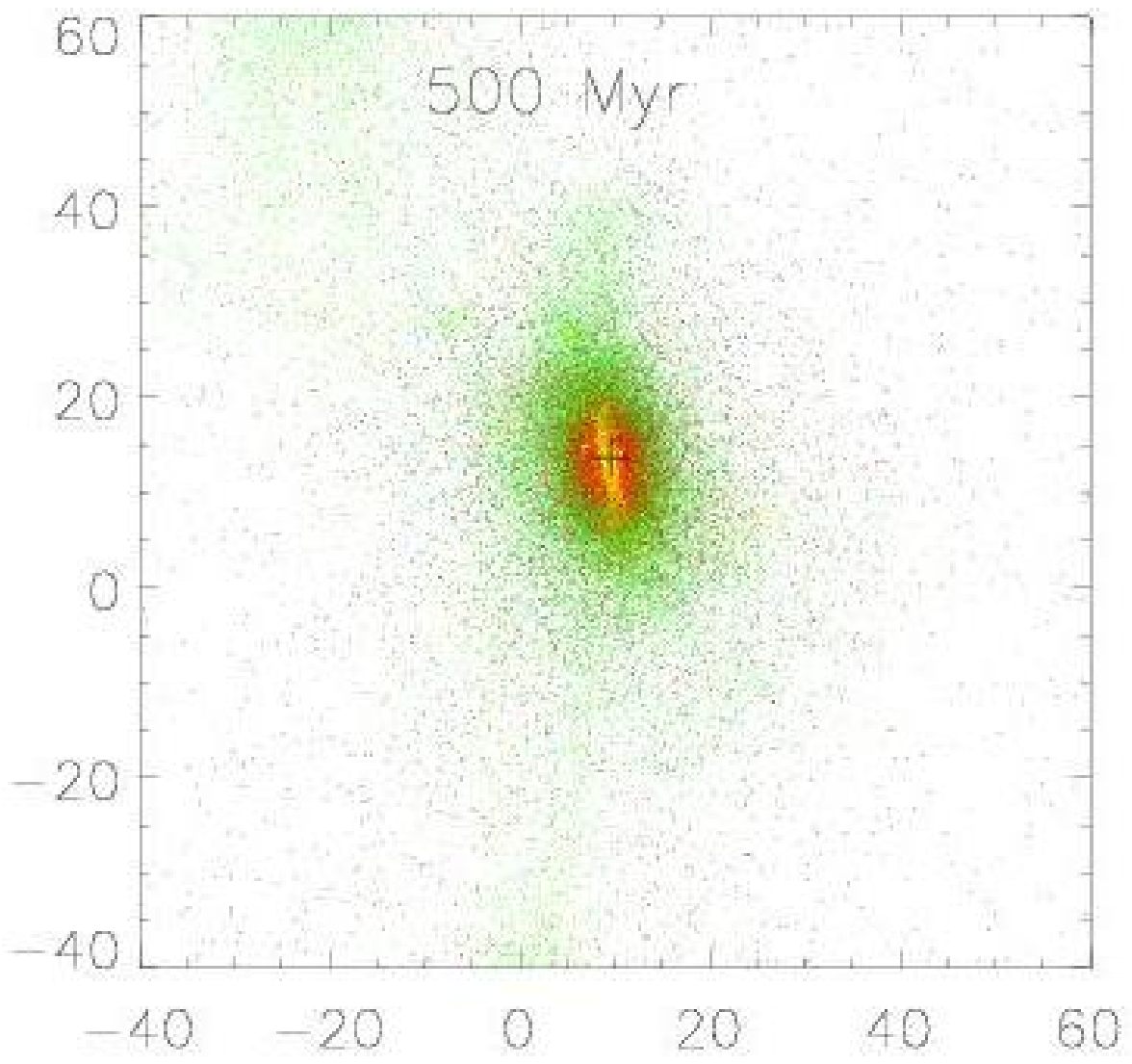,height=7.5cm,angle=0}
  }}
\end{minipage}\    \
\begin{minipage}{8.6cm}
  \centerline{\hbox{
      \psfig{figure=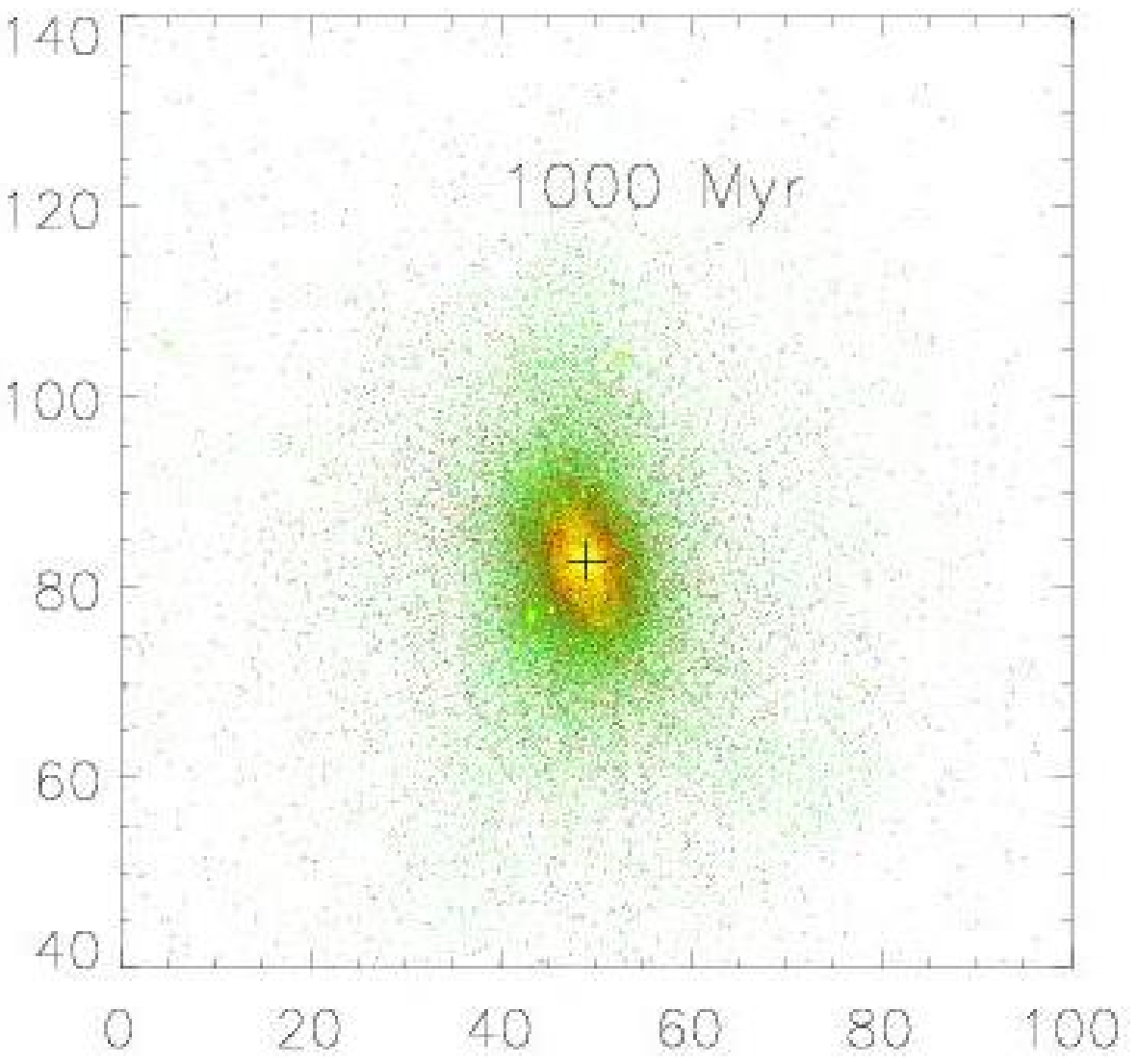,height=7.5cm,angle=0}
  }}
\end{minipage}
\caption{The same as in Fig.~1 from an orthogonal viewing angle.} 
\end{figure*}

\subsection{Modelling stellar colours with PEGASE}

Given an initial stellar mass function (for us, a Salpeter IMF), the colours of a stellar population are 
a function of how old its stars are.
We can reconstruct the colours of the hybrid particles by tracking how the gas content of each particle 
varies at each 25\,Myr.
In this way we can decompose the stars of a hybrid particles into those which formed in the last 25\,Myr, 
those which have an age $25{\rm\,Myr}<t_*<50{\rm\,Myr}$, and so on. 
If we assume that the star formation rate is uniform over intervals of 25\,Myr, then
we can reconstruct the colours of the hybrid particle with the PEGASE stellar spectral evolution model 
\citep{fioc_rocca97}. 
Since we only saved snapshots at intervals of 25\,Myr, we lack the time resolution to see nebular 
lines. This is a purely technical limitation dictated by the memory of the workstation and is easily 
overcome. Moreover, it has no consequence for the results of this paper.  

More difficult is to determine the colours of the stellar populations that preexisted the beginning of 
the simulation. For bulge stars we just assume a uniform age of 8\,Gyr at the start.
For the stellar disc we follow the method described by \citet{segalovitz75}. 
In his model the gas disc was assembled 10\,Gyr ago with no further growth from later gas cooling onto the 
disc.
The gas is in differential rotation and, owing to secular instabilities, the disc very rapidly develops 
spiral arms.
Each time a gas cloud passes through a spiral arm, a fraction $p$ of the gas  goes into stars 
and is not returned to the interstellar medium again. In this model the gas surface density evolves as 
$\Sigma_{\rm gas}(r,t)=\Sigma_{\rm gas}(r,0){\rm\,exp}(-\beta t)$. 
Here $\beta\equiv (2\pi)^{-1}n(\Omega-\Omega_{\rm p})p$,
where $n$ is the number of spiral arms, $\Omega(r)$ is the rotation angular velocity required to support the
disc and $\Omega_{\rm p}$ is the pattern angular velocity of the spiral arms. 
\citet{searle_etal73} determined the colours for a stellar population the formation of which started 
10\,Gyr ago and continued from then onward at an exponentially decaying rate on a time-scale $\beta^{-1}$. 
\citet{segalovitz75} used their determination of $B-V$ as a function of $\beta$ together with the data on 
$B-V$ as a function of galactocentric radius then available for M81. 
In this way he was able to reconstruct $\beta(r)$.
He found a good fit to the data for $p\simeq 0.2$ and $\Omega_{\rm p}\ll\Omega$. 
Here we use his model with $n=2$, $\Omega_{\rm p}=0$
and $p=0.2$ to compute $\beta(r)$ for the discs of our model galaxies. 
$\Omega(r)$ is given by the analytic formula for a
Miyamoto-Nagai density distribution \citep{miyamoto_nagai75}.
This scheme assigns an $r$-dependent age to each of the particles composing the stellar disc.
PEGASE is then used to calculate the magnitudes of these particles.
After 10\,Gyr, we expect a higher gas fraction in the outer parts of the disc, where matter has had time to
complete fewer galactic rotations. We model this effect by
attributing two different scale-radii to the gas disc and the stellar disc (Table 1).

Reality is more complicated because: i) the transformation of gas into stars is accompanied by an evolution
in the dynamic and kinematic structure of the disc, 
ii) star formation is inevitably accompanied by stellar feedback,
and iii) the outskirts of the disc will accrete fresh gas from the surrounding environment. 
However, the purpose of the model described here is simply to generate plausible initial conditions
for the colours of the merging galaxies, and this simple model accomplishes this task quite well.

\subsection{Dust absorption/emission}

The reprocessing of star light by dust is computed with the STARDUST model \citep{devriendt_etal99}. 
The optical extinction scales linearly with the column of neutral hydrogen on the line of sight and with 
the metallicity of the obscuring material. 
For a neutral hydrogen column density on the line of sight of $n_{\rm H}=10^{20}{\rm\,cm}^{-2}$ 
and solar abundances, the
$R$-, $V$- and $B$-band extinctions are $A_R\simeq 0.040\,$mag, $A_V\simeq 0.049\,$mag and $A_B=0.062\,$mag.
The chemical evolution of the interstellar medium is already included in the hydrodynamic code, 
while $n_{\rm H}$ id determined in the phase of post-processing, since it depends on the viewing angle.
All hybrid particles in front of a hybrid or stellar particle contribute towards absorbing the light of that
particle. Moreover, in the case of hybrid particles, there is also a contribution from self-absorption.  

The main difficulty derives from the fact that the SPH simulation cannot resolve the gas clouds that produce
the absorption. Hybrid particles correspond to gas clouds of $10^4 - 5\times 10^5M_\odot$, which is the 
appropriate mass range for large molecular clouds. 
If the absorption of light was uniform over a cloud's entire cross section, 
then a cloud radius of $r_{\rm cloud}$ would correspond to
$n_{\rm H}\simeq 1.3\times 10^{21}(M_{\rm cloud}/10^5M_\odot)(r_{\rm cloud}/50{\rm\,pc})^{-2}$
(assuming that 80$\%$ of the cloud mass $M_{\rm cloud}$ is composed of hydrogen). However,
in reality, molecular clouds do not absorb light uniformly because the cold gas has a fractal distribution.
Large clouds are composed of smaller clouds interspersed with voids.
Furthermore, the concept itself of associating hybrid particles with large molecular clouds is qualitatively 
useful but is dangerous when used quantitatively because
only smoothed quantities can be relied to represent physical properties.

The more sensible approach is to compute a smoothed $<n_{\rm H}>$ for the gas in front of each particle 
and then to use this $<n_{\rm H}>$ together with assumptions about the physical properties of the obscuring
clouds in order to generate a Monte Carlo $n_{\rm H}$. It is this second $n_{\rm H}$ that we eventually use 
to calculate the optical extinction.
To pass from $<n_{\rm H}>$ to a probability distribution for $n_{\rm H}$ we need to make some hypothesis 
about the column densities of the clouds that produce the extinction.
In reality, the distribution of cloud properties is very broad, but here we simply assume that all clouds 
have the same column density $n_{\rm H}^{\rm cl}$.
If $<n_{\rm H}> \ll n_{\rm H}^{\rm cl}$, then in most cases there will be no cloud or one cloud at most on
the line of sight. The probability of the second occurrence  is therefore 
$\sim <n_{\rm H}>/n_{\rm H}^{\rm cl}$,
If $<n_{\rm H}> \gg n_{\rm H}^{\rm cl}$, then typically there will be
a number of $\sim <n_{\rm H}>/n_{\rm H}^{\rm cl}$ clouds on the line of sight, and we can expect 
deviations from the average extinction to be small if this number is large.

STARDUST is used to compute the bolometric luminosity absorbed over the entire optical spectrum. 
This power is returned in the infrared as thermal dust emission with a modified blackbody spectrum.
A modified blackbody describes the physical reality better than the Planck formula because the dust is 
transparent to photons at wavelengths $\gg 200\,\mu$m, so that
radiation in that part of the spectrum is no longer thermalised.

When we compute broad-band infrared spectra for the whole galaxy, we do not
consider absorption and emission by individual clouds. Instead, we compute a smoothed 
$n_{\rm H}$ map, where one pixel corresponds to $100{\rm\,pc}\times 100{\rm\,pc}$. 
We also compute $\bar{z}_*\pm\Delta z_*$ and $\bar{z}_{\rm gas}\pm\Delta z_{\rm gas}$ at each pixel
for the distribution of stellar and hybrid particles along the line of sight ($z$ is the Cartesian
coordinate in the direction of the line of sight, not to be confused with the metallicity $Z$).
We use the results of these calculations to determine the surface brightness that is absorbed and 
reemitted. This is how we calculate the infrared emission from dust heated by star light.

We also calculate infrared spectra when the AGN contributes towards heating the dust.
We shall see later on that in our simulation $\sim 80\%$ of the lines of sight looking towards the central engine are obscured.
The obscured lines of sight have optical depth $\tau\gg 1$ (see the $n_{\rm H}$ column in Table~2).
The absorbed power is therefore an important fraction of the AGN's bolometric luminosity.
Therefore, we can make the approximation that there is an optically thick central region, which absorbs $\sim 80\%$ of
the AGN power and reemits it with the modified black body spectrum described above. 
The radius of the self-absorbed central region gives
the characteristic radius of the surface that reemits the absorbed AGN light.
We compute this radius to estimate the temperature of the reemitted radiation.

\subsubsection{Creating mock images}

The software package SkyMaker (http://terapix.iap.fr/soft) is used to convert the projected coordinates and 
extinction-corrected 
magnitudes of the simulated particles
into mock images in the Johnson $B$, Johnson $V$ and Cousins $R$ bands. These images are generated as files
in the .fits format. 
Therefore, it is possible to perform on them all the manipulations that could be performed on real data. 
SkyMaker can model instrumental diffraction, photon noise and atmospheric blurring.
Since it is possible to put the galaxies at an arbitrary distance, this technology
can be used not only to explore theoretical scenarios for galaxies and AGN, but also to assist the 
development of new observing proposal by determining in advance which features would be observable at a 
given flux limit or under certain sky conditions (in the case of ground-based telescopes).

Fig.~3 shows a true colour image of one of the two galaxies before the merger takes place.
This snapshot is generated by using the RVB colour system to superimpose images in the $B$, $V$ and $R$ bands.
This image shows the dust lane cutting through the galactic disc.
\begin{figure}
\begin{center}
\centerline{\psfig{figure=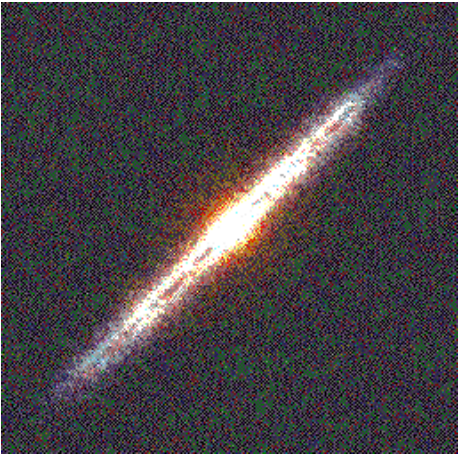,height=6.cm,angle=0}}
\end{center}
\caption{A close up image of galaxy 1 viewed edge-on shows the central dust lane.}
\end{figure}

\section{The black hole}

\subsection{The approach}

We assume that the two galaxies start with central black holes of $3\times 10^7M_\odot$. 
This mass is inferred from the black hole mass to bulge mass mass relation in \citet{haering_rix04}.

The mass resolution of the simulation is not sufficient to compute the orbits of the black holes
correctly because, while in reality 
the black hole mass is much larger than the stellar mass, so that the black holes sink to the centre due to dynamical friction, 
that is not the case in our simulation.
Therefore, we impose as a constraint that the black hole is in the galactic nucleus at all time 
and calculate the accretion rate from the fuel supply that is present in the nuclear region.
This approach relies on a correct identification the galactic centre.

\subsection{Identifying the galactic centre}

The identification of density peaks in the simulation outputs is done with the
substructure finder adaptaHOP,
described in detail in \citet{Aubert_etal03}. Basically, this
algorithm applies SPH smoothing to the particle distribution,
then finds local
density maxima and saddle points in this smoothed distribution and
constructs a tree of structures and substructures by using the value of the density at
saddle points to control connectivity between various subcomponents of the system. 
The local maxima correspond to the leaves of the tree. 
A given substructure is identified as the set of particles verifying $\rho_{\rm SPH} \geq
\rho_{\rm saddle}$, where $\rho_{\rm SPH}$ is the SPH density associated
to each particle and $\rho_{\rm saddle}$
is the value of the SPH density at the (highest) saddle point connecting
this substructure to its neighbour.
The output of adaptaHOP depends
on the choice of various parameters, namely the number $N_{\rm SPH}$ of
particles used to perform SPH smoothing,
the number $N_{\rm HOP}$ of neighbours used to walk into the particle
distribution in order to
find local maxima and to establish connectivity through saddle points, and
a parameter $f_{\rm Poisson}$
controlling the significance of substructures compared to Poisson
noise. 
Our choice for these parameters is the one advocated in
\citet{Aubert_etal03}, namely $N_{\rm SPH}=64$,
$N_{\rm HOP}=16$ and $f_{\rm Poisson}=4$, the latter value meaning that
only substructures at 4$\sigma$ level compared
to pure Poisson noise are considered as real. Besides these parameters,
there is as also an overall density threshold, $\rho_{\rm TH}$, which is used to
eliminate all particles with SPH density below $\rho_{\rm TH}$ prior to the analysis.
That we set here to a very small value (in cosmological simulations, $\rho_{\rm TH}$
controls the selection of dark matter haloes).

In adaptaHOP, a galaxy is thus viewed as a tree, which separates in
several branches, which separate into
smaller branches, which separate into leaves, corresponding to small scale
density peaks.
This tree represents how smaller structures are nested into larger
structures through saddle points
at a fixed time and has nothing to do with merger trees encountered
in cosmological simulations of galaxy formation.
We search for density peaks in the leaves of the tree by starting from the
leaves that contain the highest
masses. This usually gives a good identification of the galactic nucleus.
Instead, by starting from the leaves with the highest density one often
misses the galactic nucleus and
finds molecular clouds in the spiral arms.
One can look at Fig.~1 and see that the centres identified by this
automatic procedure,
marked by little black crosses,  are, as they should be, at the centres of
the two galaxies.
Identifications are made secure by the fact that the galactic nucleus is
not just a name for the central
region, but stands out as a well defined dynamical entity (see the right
hand column of Fig.~5).
Moreover, we have verified that the particles which are in the galactic
nucleus (identified as the
region at a radius of $r<100{\rm\,pc}$) at the beginning of the
simulation, remain in the galactic nucleus
during the entire simulation.
A small fraction are stripped from the galactic nucleus in mini tidal
tails,
but they eventually fall back into it.

\begin{figure*}
\noindent
\begin{minipage}{8.cm}
  \centerline{\hbox{
      \psfig{figure=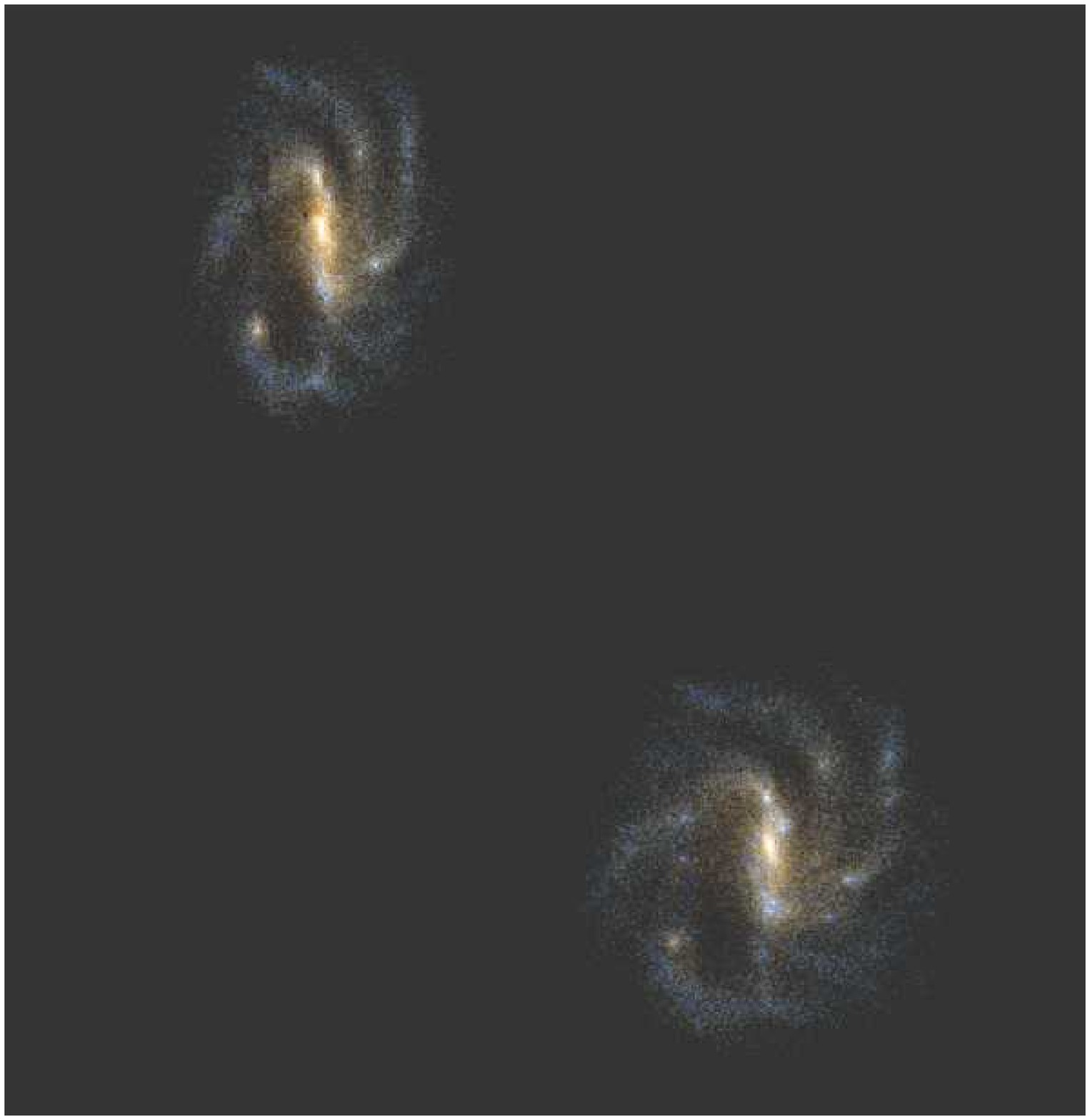,height=5.4cm,angle=0}
  }}
\end{minipage}\    \
\begin{minipage}{8.cm}
  \centerline{\hbox{
      \psfig{figure=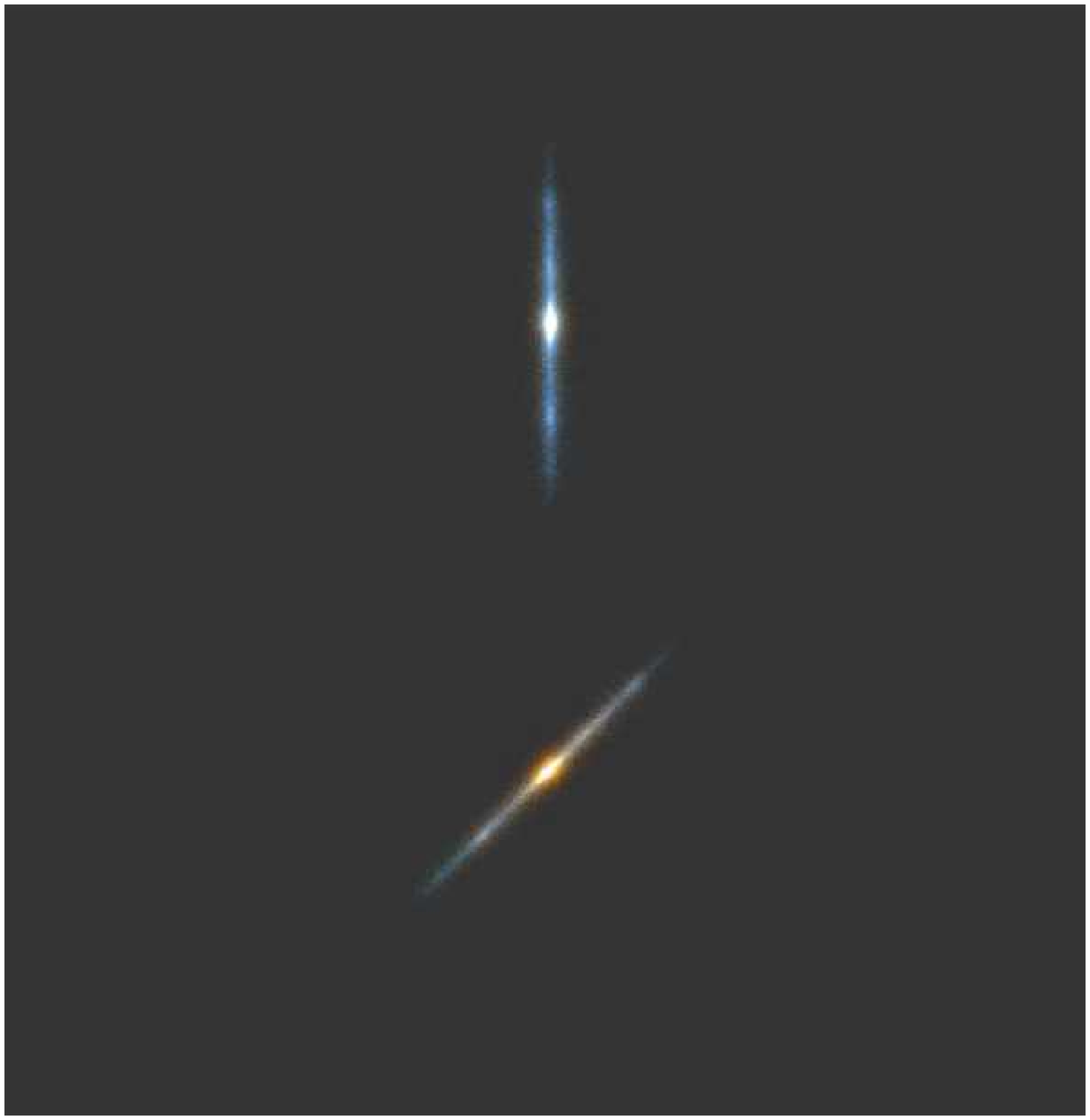,height=5.4cm,angle=0}
  }}
\end{minipage}\    \
\begin{minipage}{8.cm}
  \centerline{\hbox{
      \psfig{figure=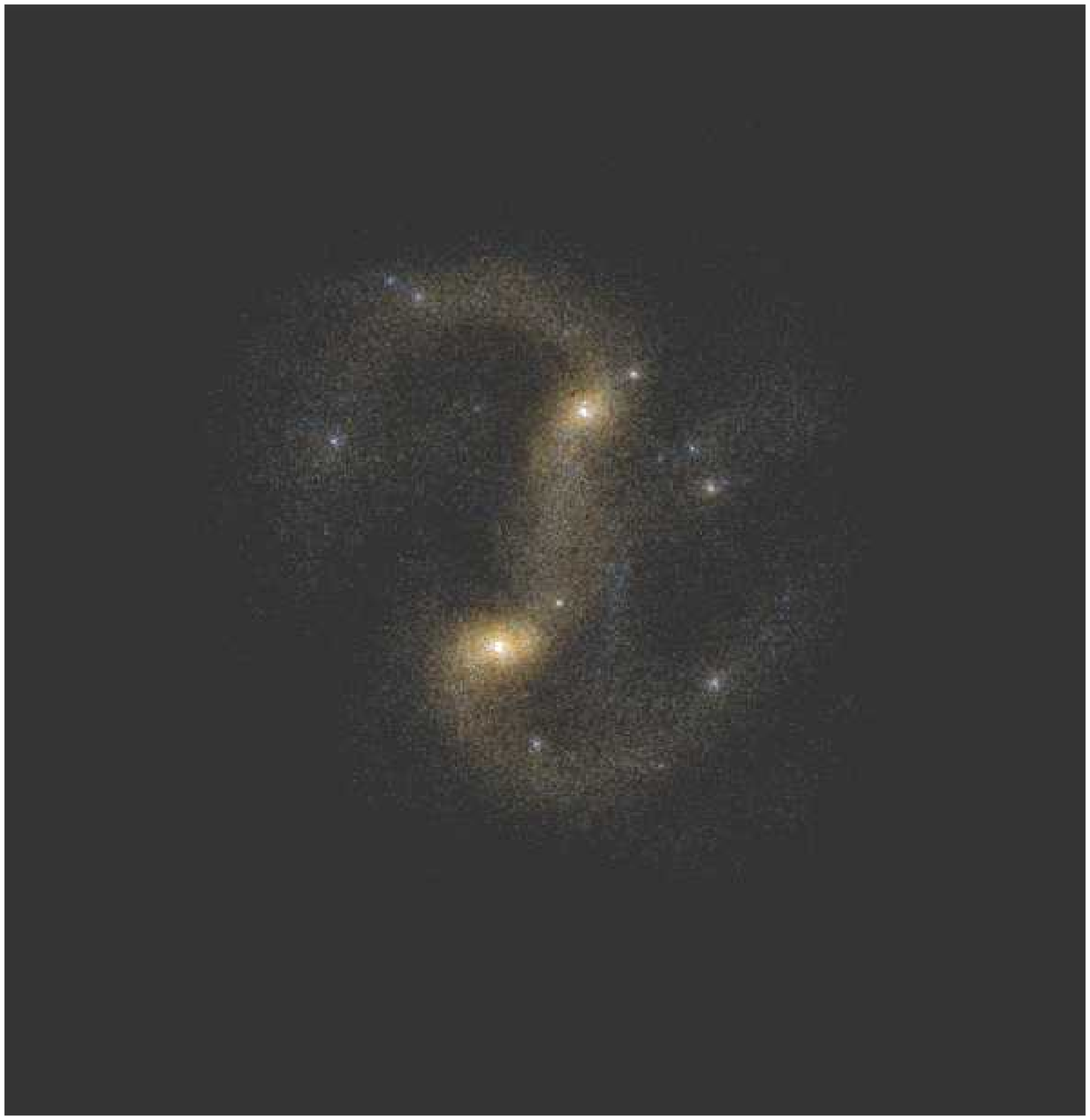,height=5.4cm,angle=0}
  }}
\end{minipage}\    \
\begin{minipage}{8.cm}
  \centerline{\hbox{
      \psfig{figure=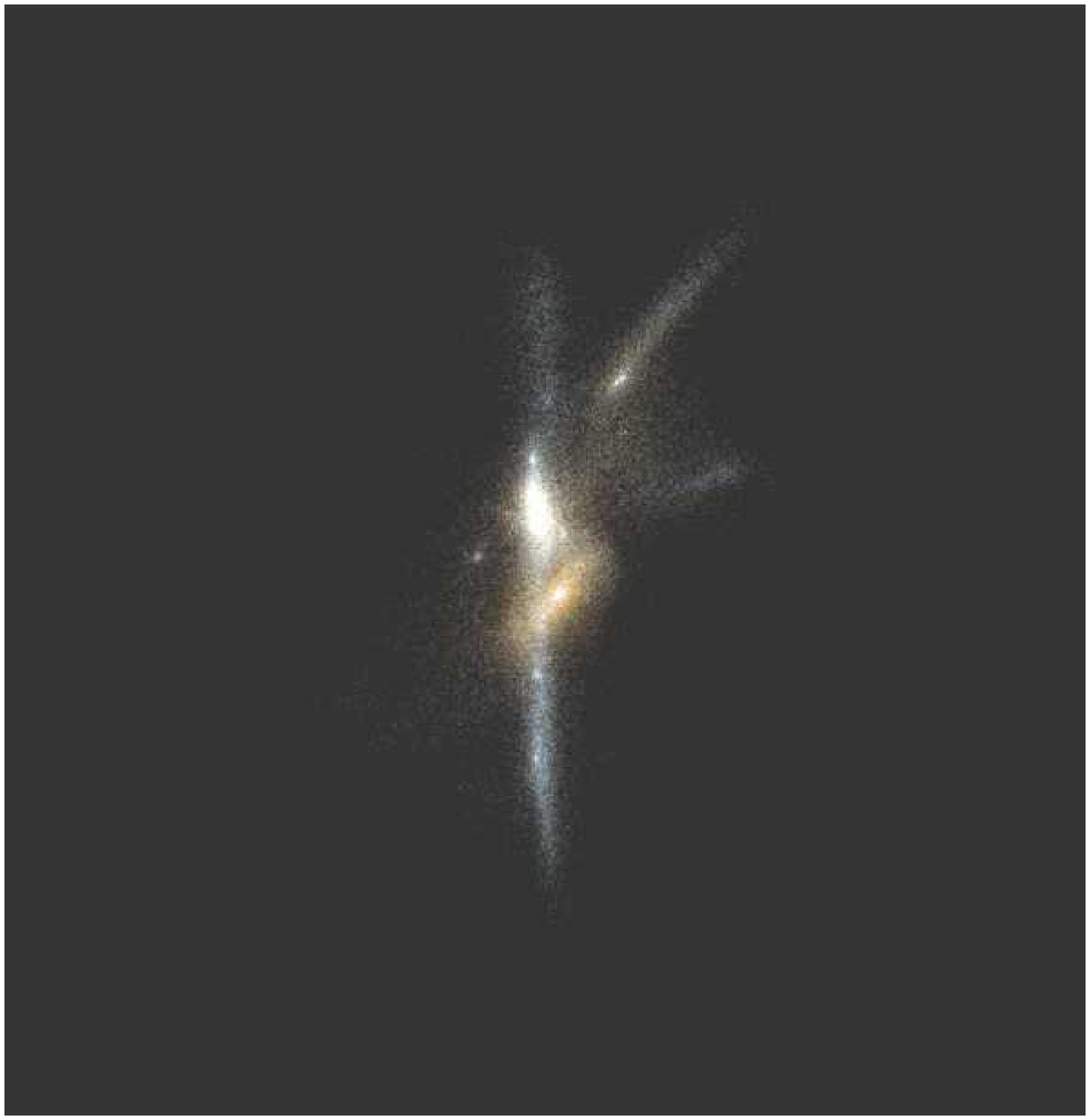,height=5.4cm,angle=0}
  }}
\end{minipage}\    \
\begin{minipage}{8.cm}
  \centerline{\hbox{
      \psfig{figure=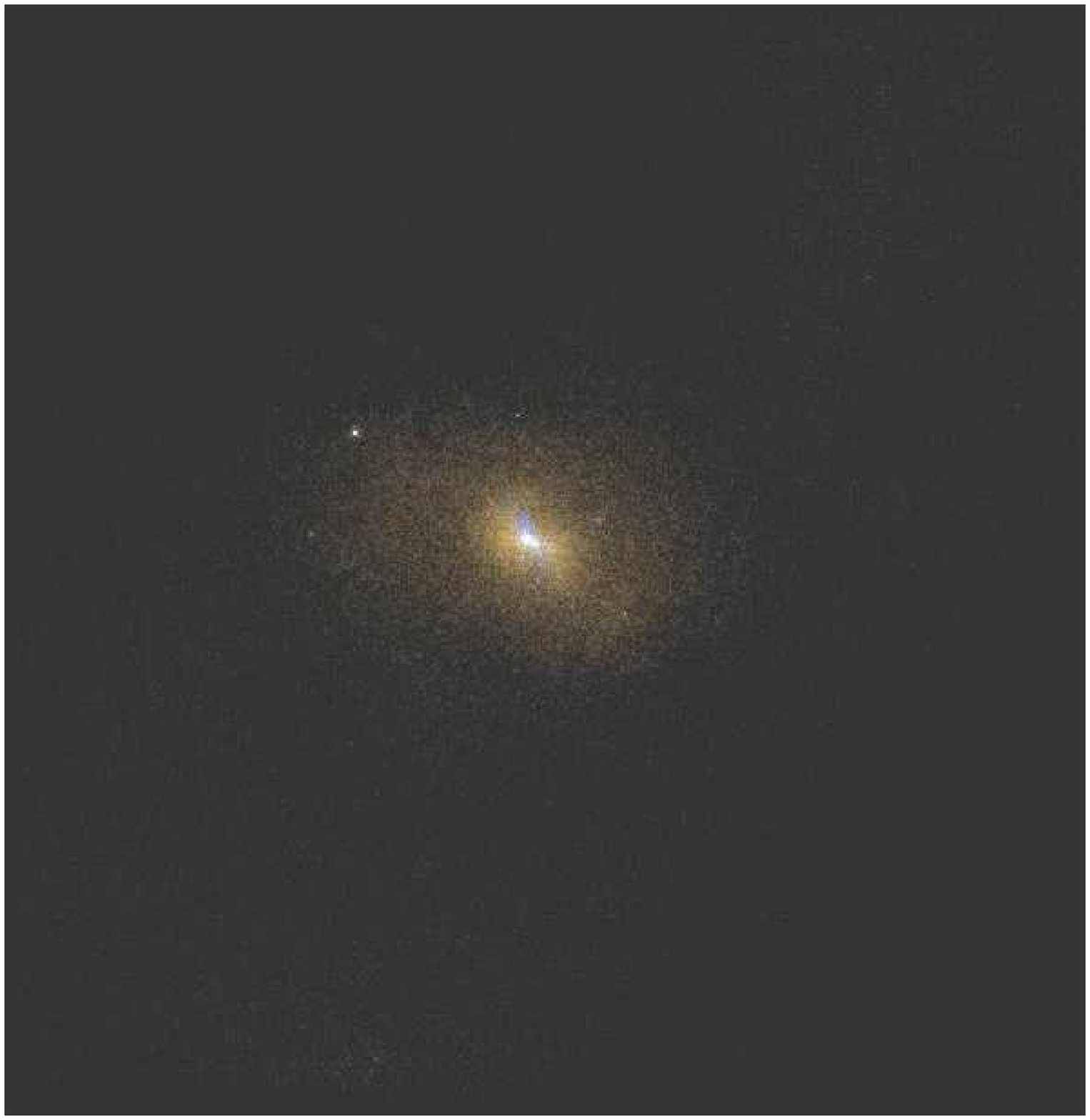,height=5.4cm,angle=0}
  }}
\end{minipage}\    \
\begin{minipage}{8.cm}
  \centerline{\hbox{
      \psfig{figure=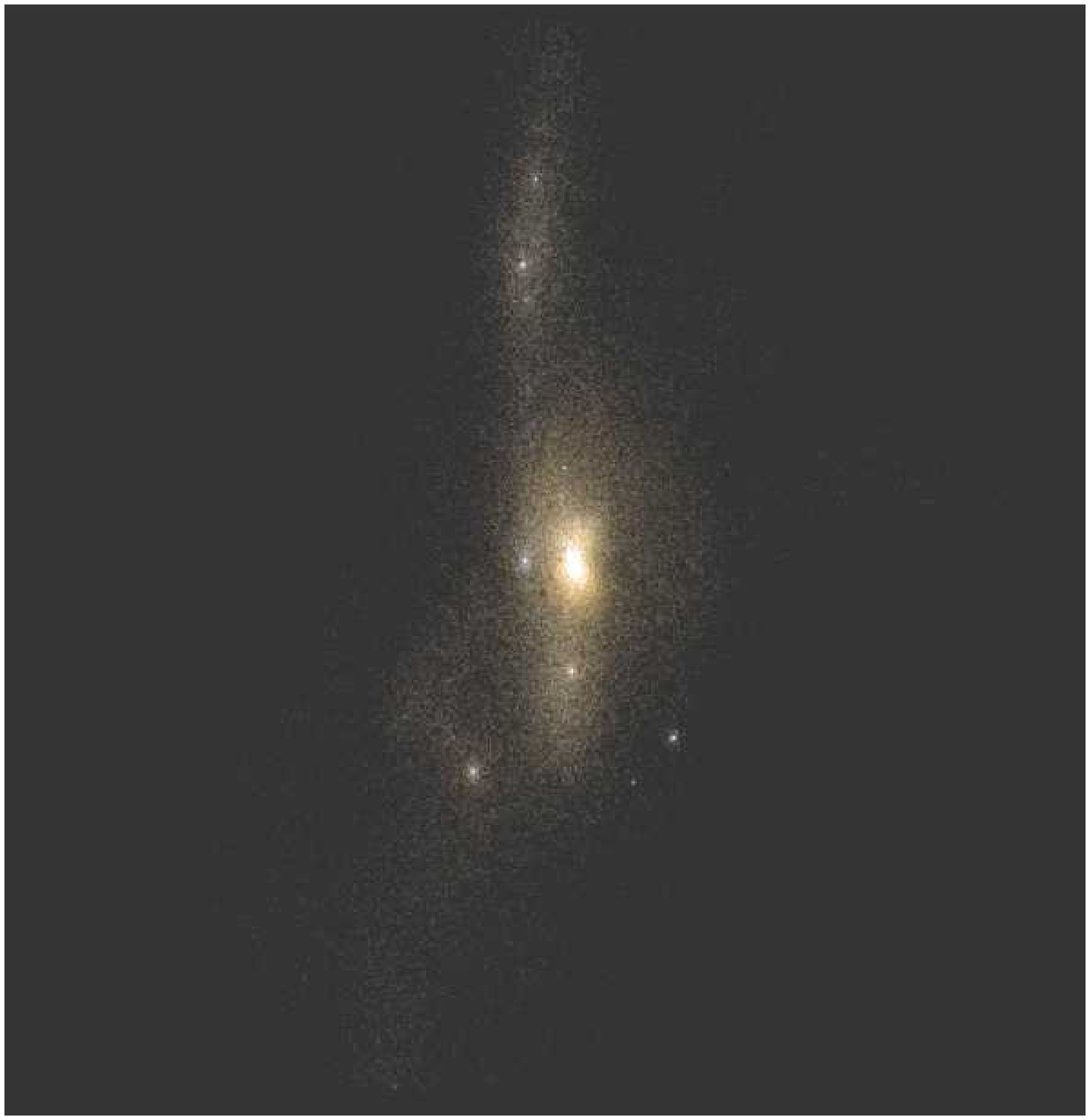,height=5.4cm,angle=0}
  }}
\end{minipage}\    \
\begin{minipage}{8.cm}
  \centerline{\hbox{
      \psfig{figure=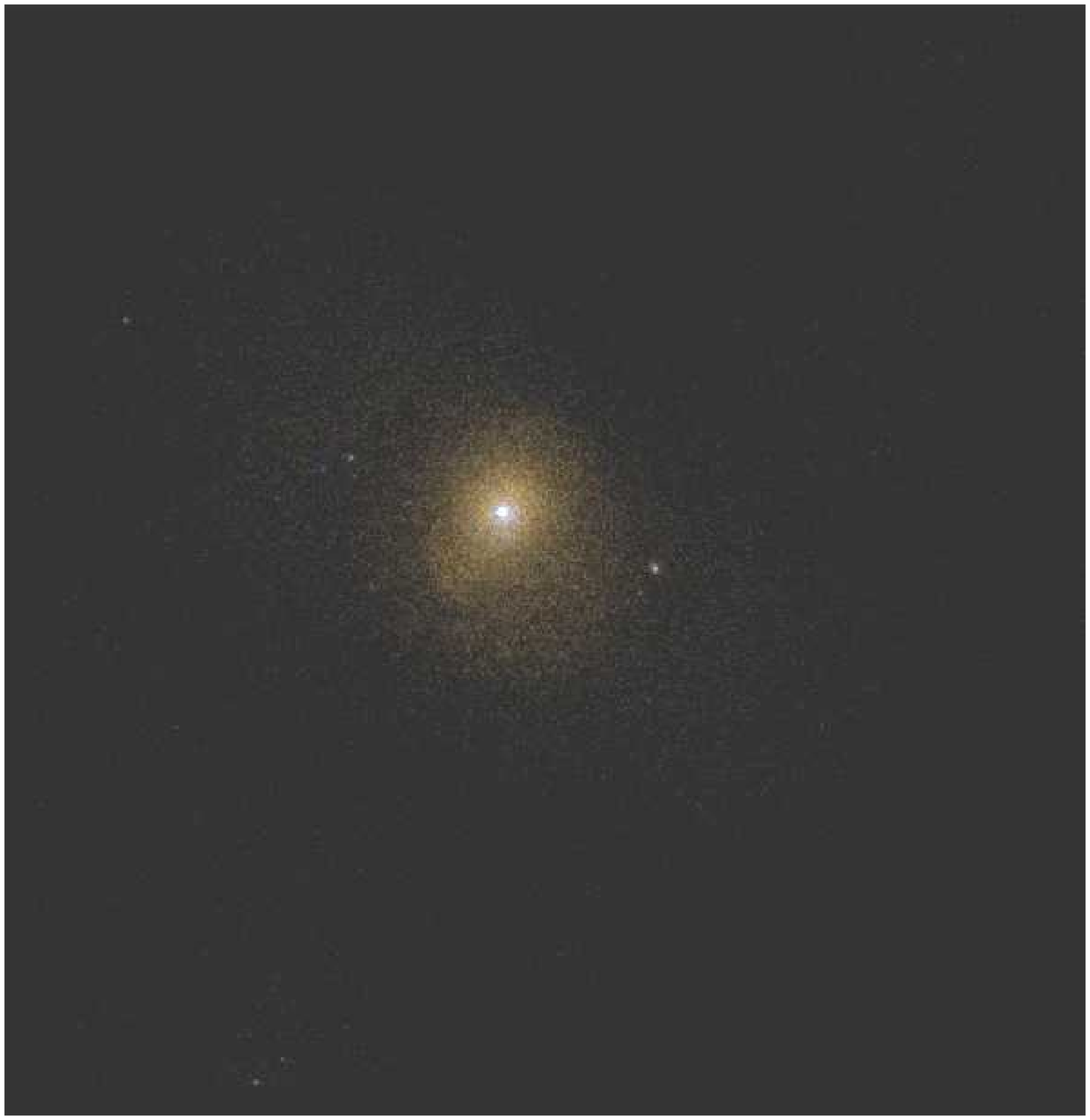,height=5.4cm,angle=0}
  }}
\end{minipage}\    \
\begin{minipage}{8.cm}
  \centerline{\hbox{
      \psfig{figure=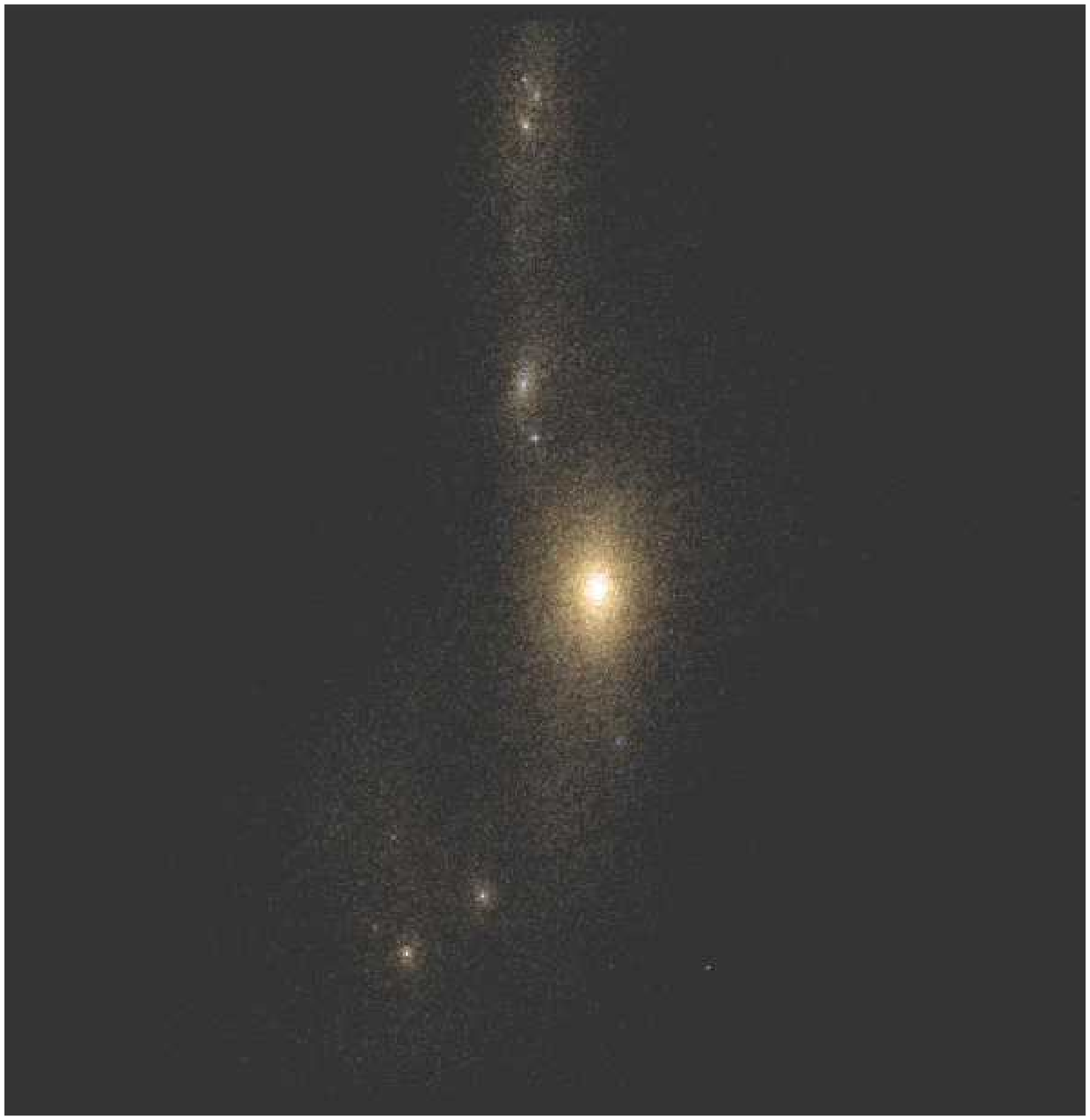,height=5.4cm,angle=0}
  }}
\end{minipage}
\caption{Face-on (left column) and edge-on (right column) snapshots of the merging process.
From top to bottom, the snapshots were taken 100, 200, 300 and 350\,Myr after the simulation started.
The snapshots were generated with the SkyMaker software by superimposing B, V and R band images.
Each image corresponds to a square of $80\,$kpc side.}
\end{figure*}

\begin{figure*}
\noindent
\begin{minipage}{8.cm}
  \centerline{\hbox{
      \psfig{figure=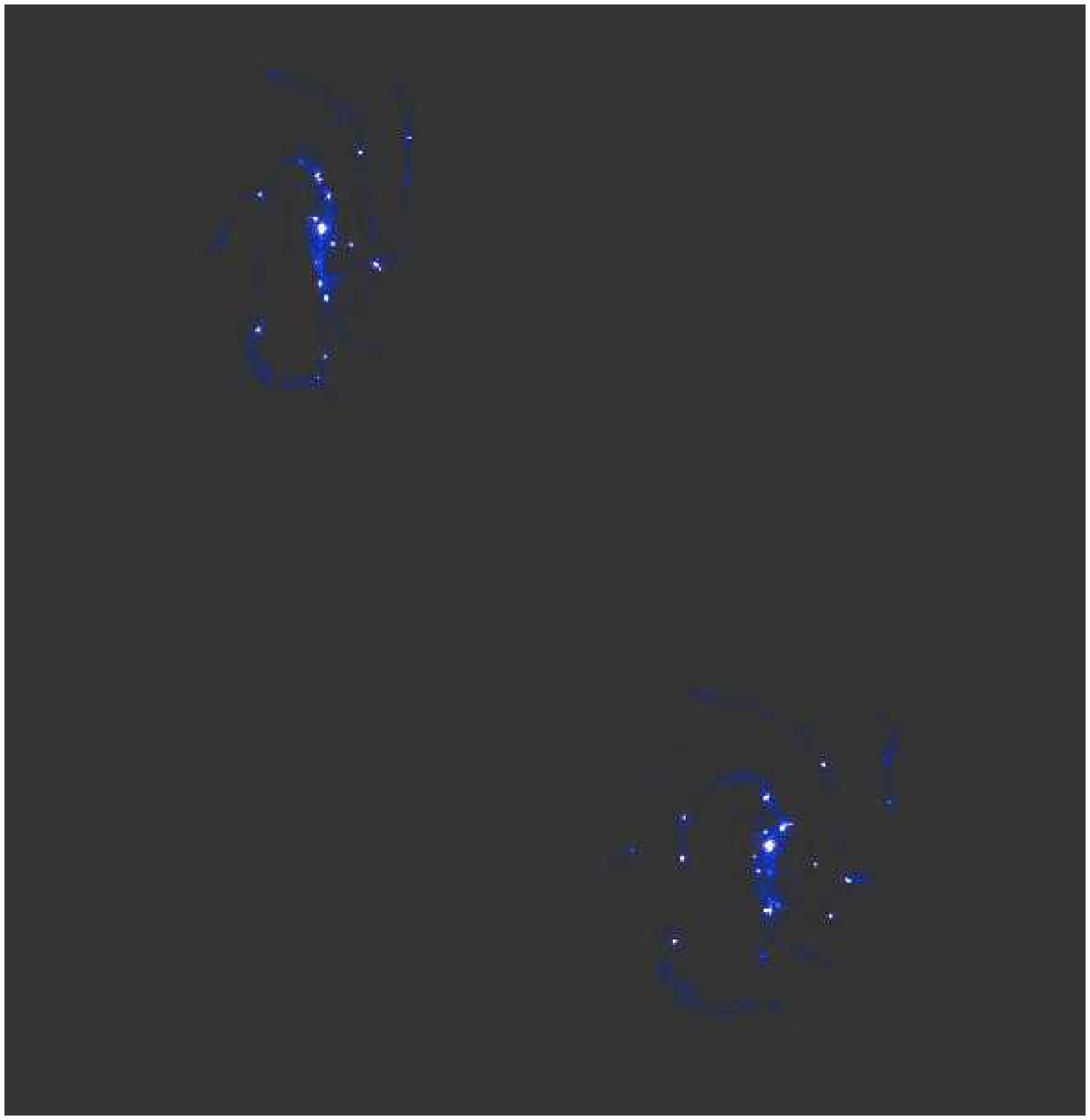,height=5.4cm,angle=0}
  }}
\end{minipage}\    \
\begin{minipage}{8.cm}
  \centerline{\hbox{
      \psfig{figure=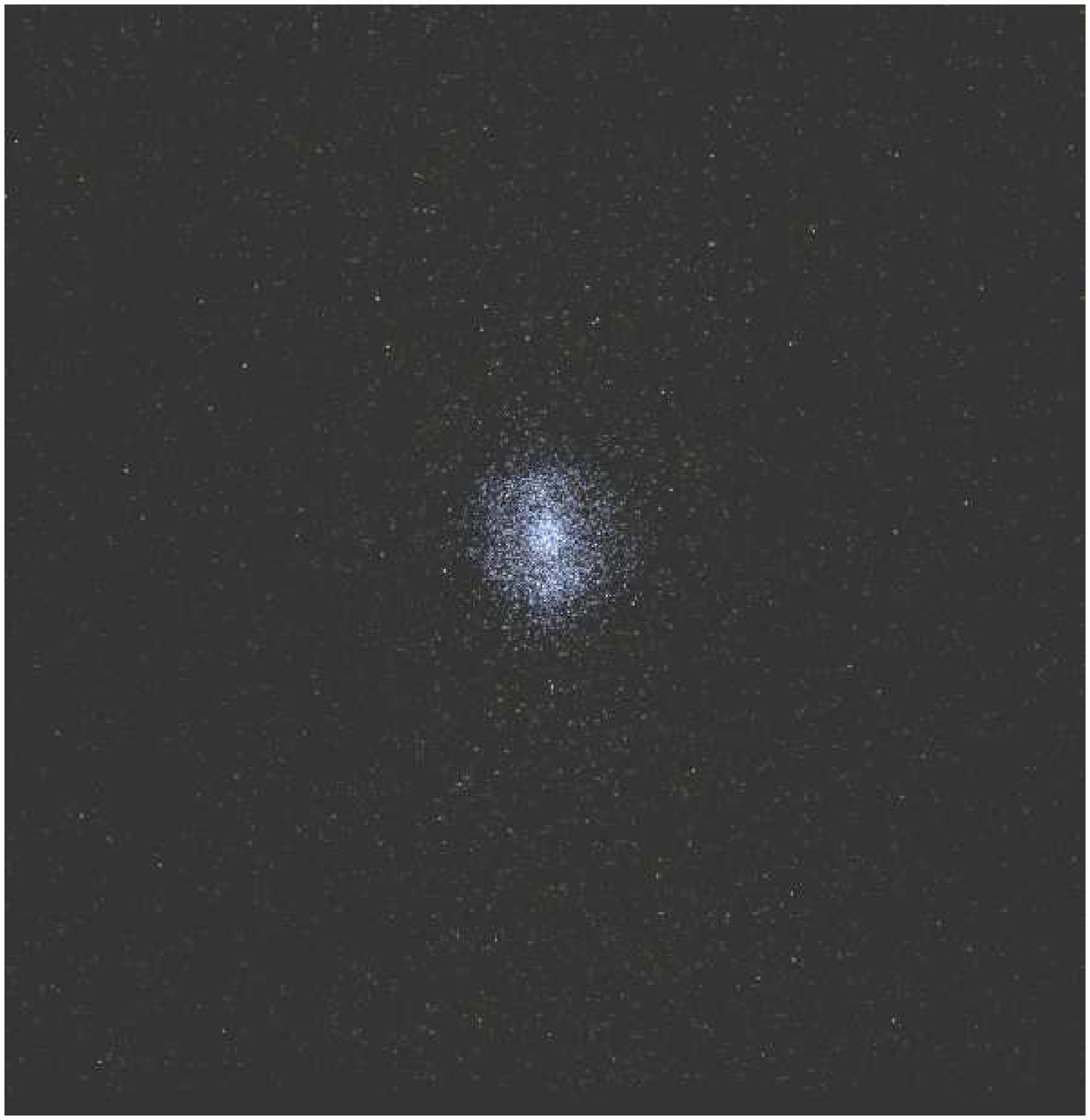,height=5.4cm,angle=0}
  }}
\end{minipage}\    \
\begin{minipage}{8.cm}
  \centerline{\hbox{
      \psfig{figure=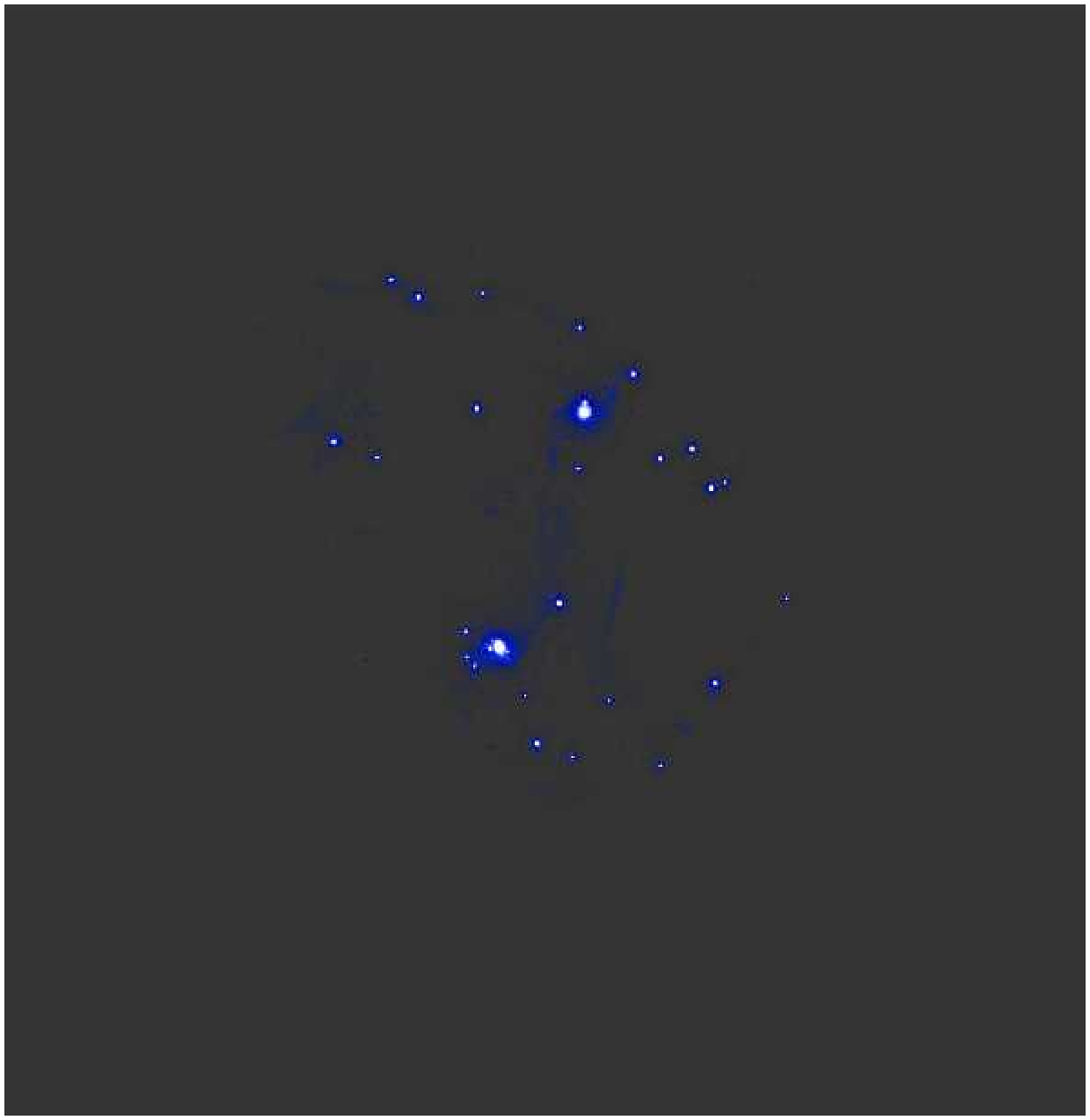,height=5.4cm,angle=0}
  }}
\end{minipage}\    \
\begin{minipage}{8.cm}
  \centerline{\hbox{
      \psfig{figure=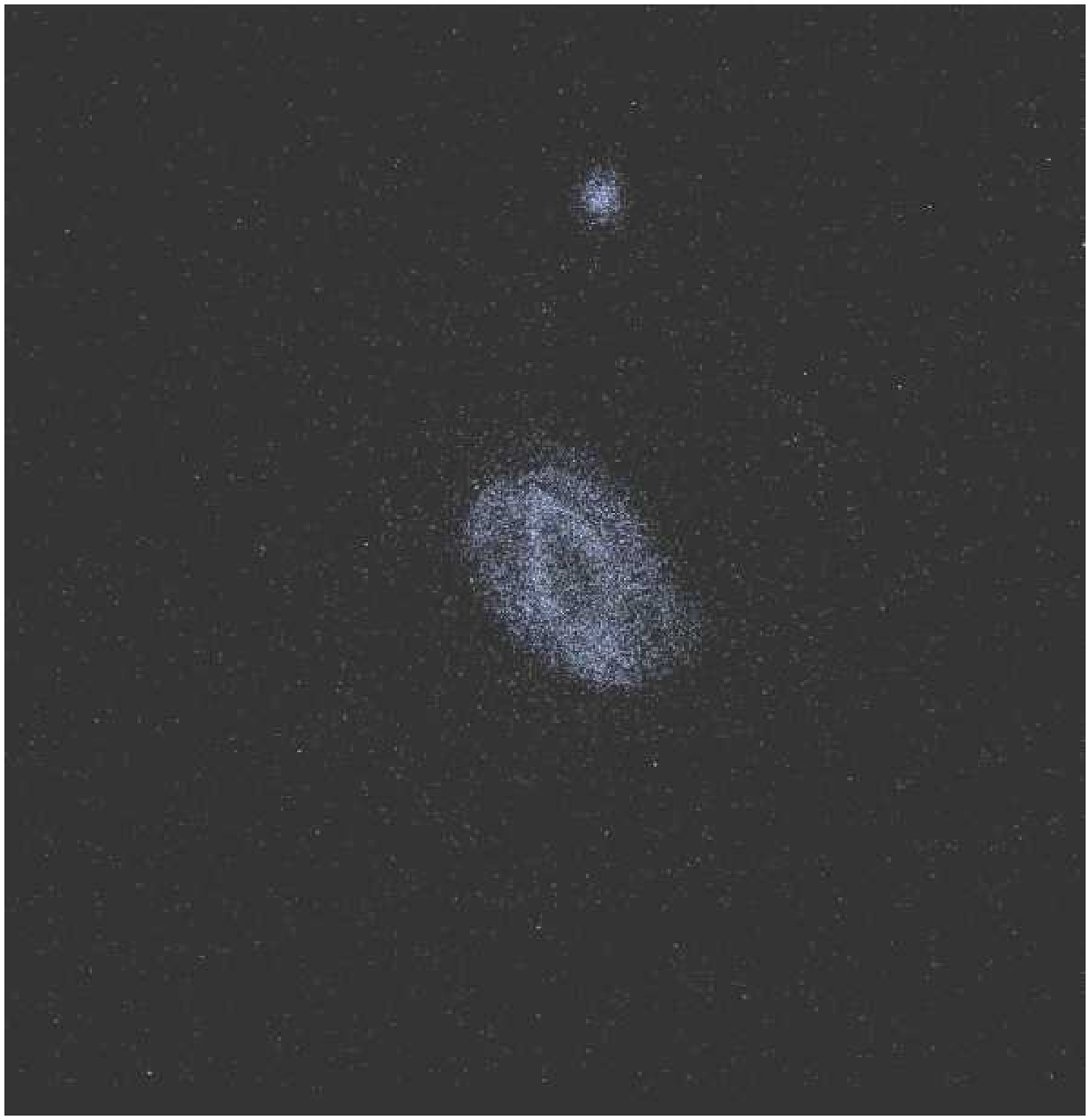,height=5.4cm,angle=0}
  }}
\end{minipage}\    \
\begin{minipage}{8.cm}
  \centerline{\hbox{
      \psfig{figure=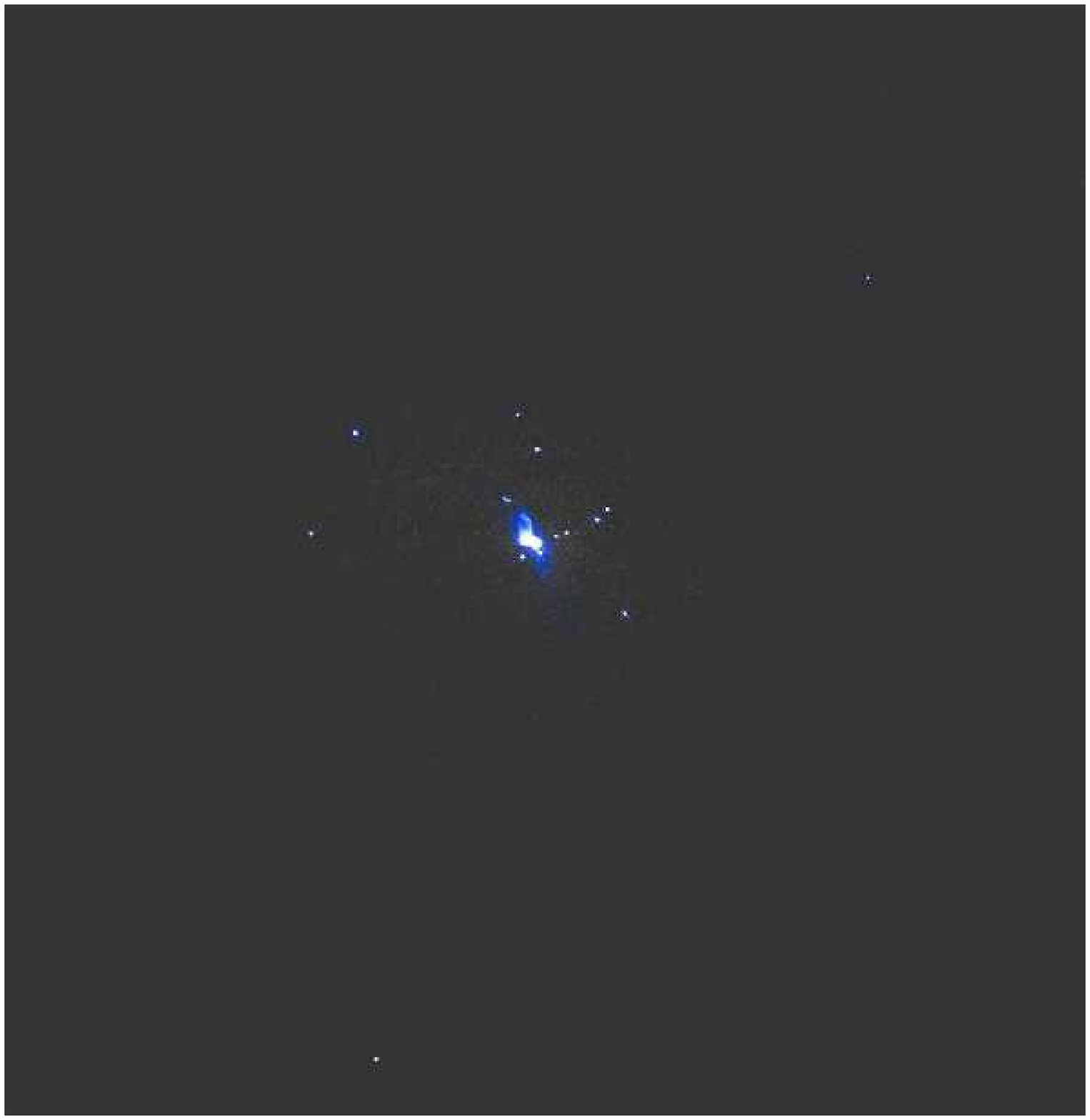,height=5.4cm,angle=0}
  }}
\end{minipage}\    \
\begin{minipage}{8.cm}
  \centerline{\hbox{
      \psfig{figure=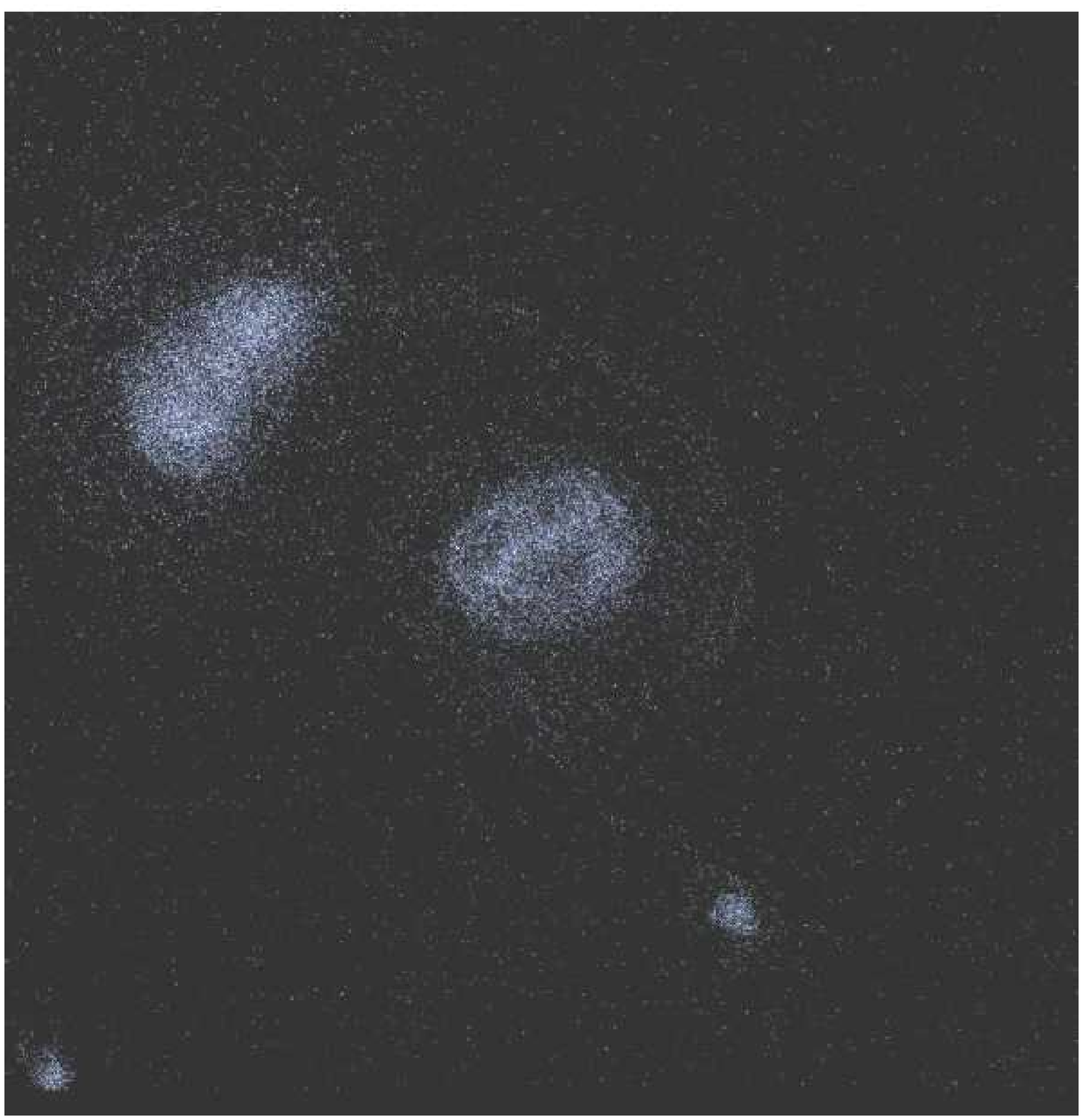,height=5.4cm,angle=0}
  }}
\end{minipage}\    \
\begin{minipage}{8.cm}
  \centerline{\hbox{
      \psfig{figure=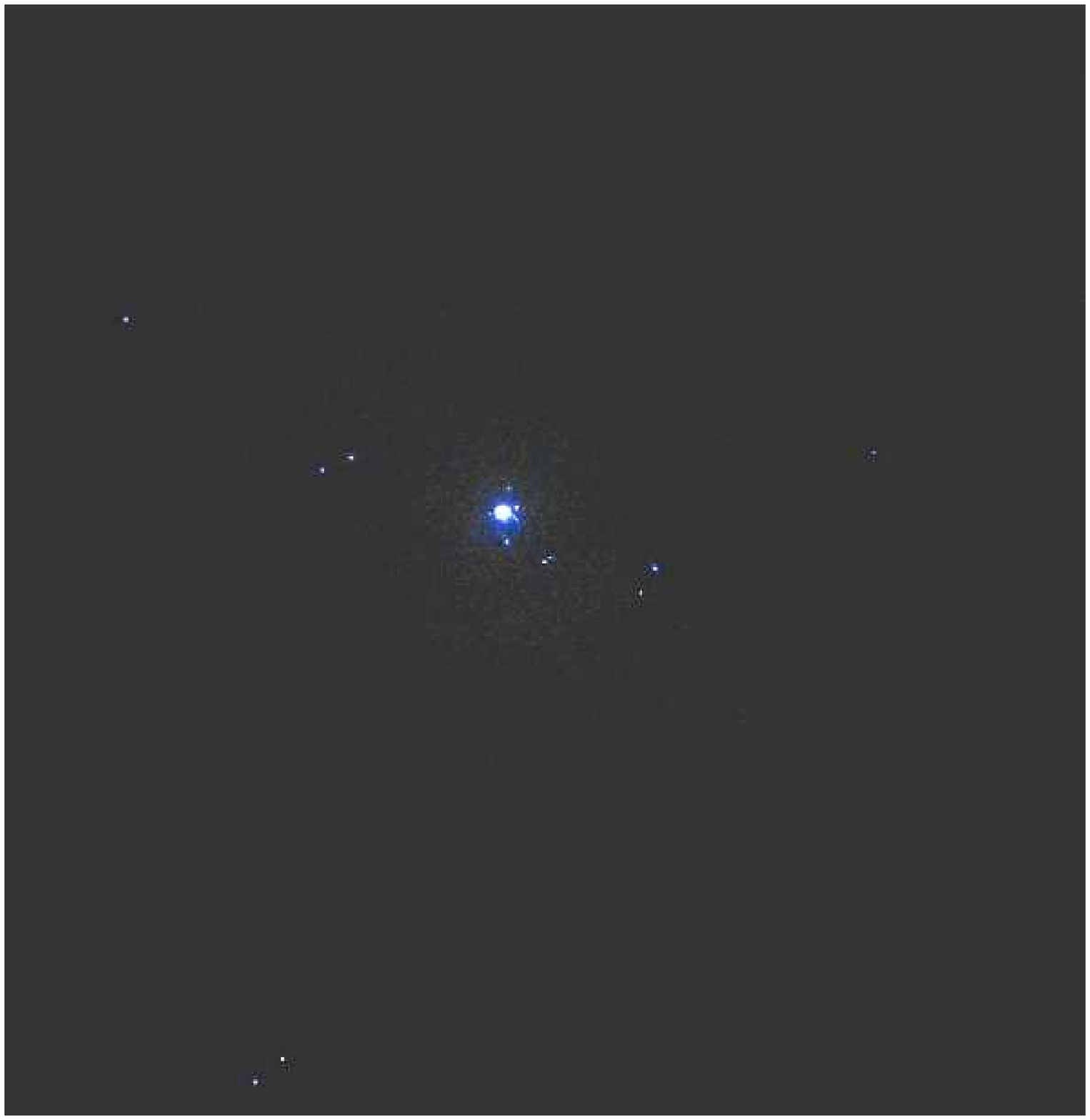,height=5.4cm,angle=0}
  }}
\end{minipage}\    \
\begin{minipage}{8.cm}
  \centerline{\hbox{
      \psfig{figure=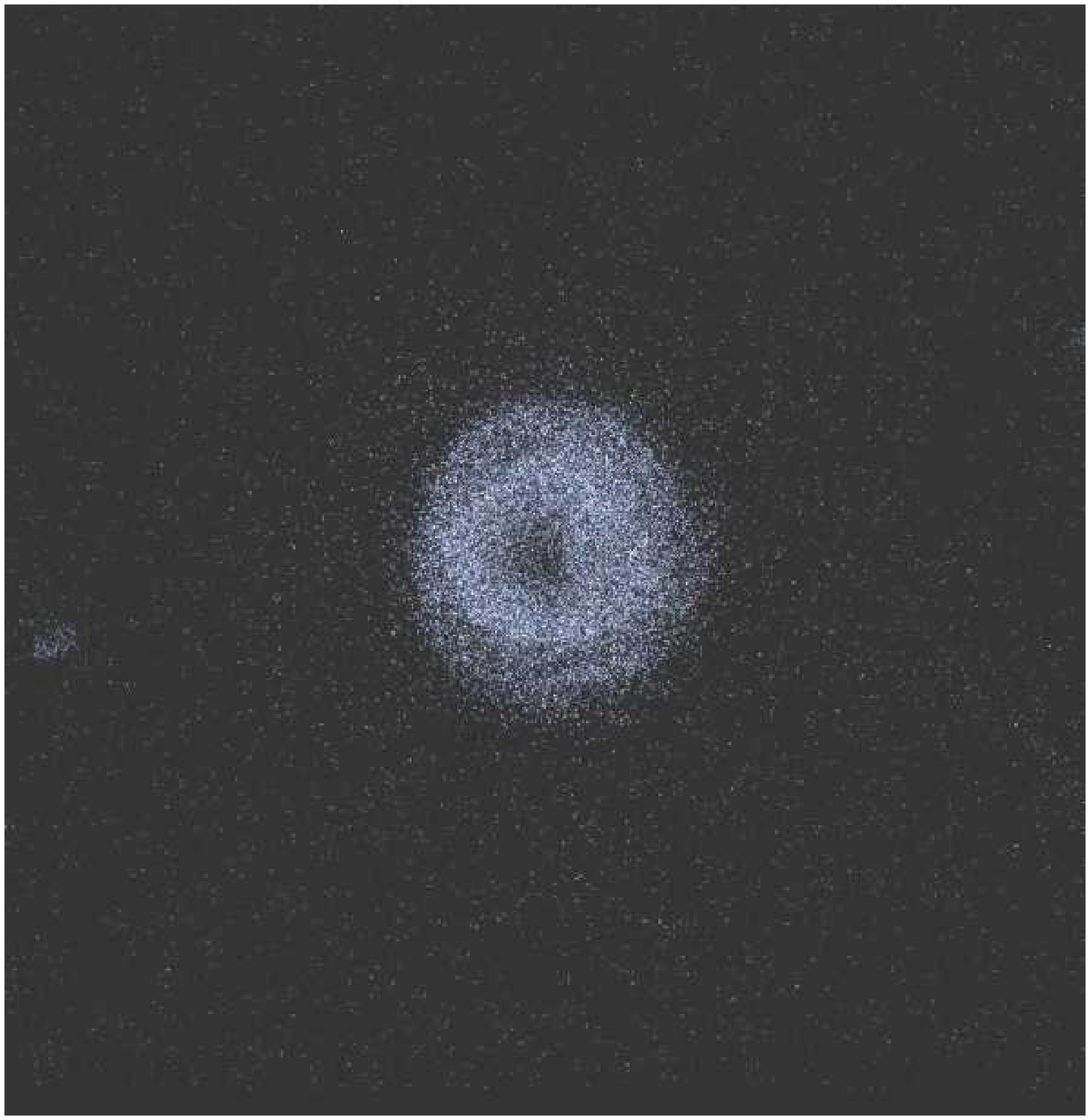,height=5.4cm,angle=0}
  }}
\end{minipage}
\caption{The left column corresponds to left column of Fig.~4 with the difference that here we only
show the young stellar population formed after the simulation began. In these four snapshots, 
and in these four snapshots only, we have neglected the extinction of starlight by dust.
The right column shows a $2{\rm\,kpc}\times 2{\rm\,kpc}$ zoom of the nucleus of galaxy 1.
These snapshots were derived by zooming into the images on the left column of Fig.~4.
Young stars dominate the emission.}
\end{figure*}

\subsection{Fuelling the central black hole by orbital decay of large molecular clouds}

Our SPH simulation can follow the inflow of gas from a few kpc into the central 100$\,$pc, 
but lacks the resolution to follow the inflow of gas from a 100\,pc distance into the central black hole. 
Following \citet{shlosman_noguchi93}, we imagine that the stream of matter from the galactic 
nucleus into the black hole is due to the orbital decay of giant molecular clouds, which are subject to the 
dynamical friction of stars and of lower mass clouds and which are eventually disrupted by tidal forces. 
At that point the clouds are stretched into a continuous accretion flow into the black hole. 
The formula for the accretion rate from the orbital decay of molecular cloud is:
$$\dot{M}_\bullet=2\times 10^{-3}{\psi\over 0.1}{M_{\rm cloud}\over 10^5M_\odot}
{M_{\rm gas}\over 10^9M_\odot}
{\rho_*\over M_\odot{\rm\,pc}^{-3}}\times$$
\begin{equation}
\times {v_{\rm a}\over 10{\rm\,km\,s}^{-1}}\left({v_{\rm c}\over 100{\rm\,km\,s}^{-1}}\right)^{-1}
\left({\sigma_*\over 25{\rm\,km\,s}^{-1}}\right)^{-3}{M_\odot\over{\rm yr}}
\end{equation} 
\citep{shlosman_noguchi93}, 
where $\psi$ is the fraction of the total gas in the galactic nucleus $M_{\rm gas}$ that is contained in
clouds of mass $>M_{\rm cloud}$,
$\rho_*$ is the stellar density in the galactic nucleus, 
$\sigma_*$ is the stellar velocity dispersion,
$v_{\rm a}$ is the relative velocity at which the gas rotates with respect to the stars  
and $v_{\rm c}$ is the rotational velocity of the gas.
We extract all these quantities directly from the outputs of the SPH simulation and use them to compute
$\dot{M}_\bullet$. The two black holes merge when we can no longer separate the nuclei of the two galaxies.
The reader should be warned that the precise value and evolution of
$\dot{M}_\bullet(t)$ is sensitive to the details of how we calculate the average quantities mentioned above,
but we have verified that none of our conclusions depends on these details.
Moreover, our model for the black hole accretion rate tends to smooth the accretion rate over time.
The real pattern of black hole accretion is more likely to be a sequence of bursts with variability
on all time-scales. The most pragmatic approach to deal with this reality when necessary is to
introduce an AGN duty cycle, so that the quasar is only active for a fraction of the simulation time.
The accretion rate during the active phase is correspondingly increased to account for the presence
of this duty cycle.

\section{Results}

The evolution of the system during the merger is shown by the sequence of snapshots in Figs.~4-5.
For each of the four represented time-steps, which correspond to 100, 200, 300 and 350\,Myr from the
beginning of the simulation, we show an $80{\rm\,kpc}\times 80{\rm\,kpc}$ face-on image 
(Fig.~4, left column), an $80{\rm\,kpc}\times 80{\rm\,kpc}$ edge-on image (Fig.~4, right column), 
an $80{\rm\,kpc}\times 80{\rm\,kpc}$ face-on image, where we only show the unextinguished young 
stellar population (the light coming from hybrid particles, Fig.~5, left column),
and a $2{\rm\,kpc}\times 2{\rm\,kpc}$ zoom of the nuclear region of galaxy 1 (Fig.~5, right column).
See the end of Section 2 for an explanation of which one is galaxy 1.
Enlarged images of the central region are taken from face-on images.
Compare these images with Figs.~1-2 and
notice that SkyMaker reverses the original images.

On a large scale, the two galaxies are merging in a dynamical time. Their relative orbit reaches a point
of closest approach and has only time to make a turn before the galaxies merge owing to dynamical friction.
On a small scale, inside each galaxy, the tidal forces trigger matter inflow.
This process culminates with the merger of the two nuclei, which occurs about $50\,$Myr after the 
galaxies as a whole have merged, and ends with a stream of debris that keeps on falling into the nuclear
region during the post-merger phase of dynamical relaxation.

\subsection{Star forming regions}
The most obvious result coming out of these images is that strong star formation is concentrated in a
series of knots along gas spiral arms, which are much more pronounced and filamentary than the spiral 
arms in the stellar disc. The galactic nucleus stands out as the central one and the most prominent of
these knots. As the simulation progresses, these knots, associated to the formation of massive star
clusters, fall to the centre and merge with the galactic nucleus.

The star formation rate in the host galaxy is a good tracer of the star formation rate in the nuclear 
region (Fig.~6). For this reason, we do not find evidence for a delay between star formation in the host
galaxy and supply of fuel to the central region.

The first peak of star formation in Fig.~6 is exaggerated by the artificial assumption that the discs 
are initially axisymmetric (Fig.~7). The assumed initial condition is highly unstable.
The first peak of star formation would have been absent if we had started from a relaxed dynamical configuration.
The second peak corresponds to a physical phenomenon, the starburst triggered by the merger.
During this starburst the star formation rate is about six times higher than it is in an isolated galaxy.

\begin{figure}
\psfig{figure=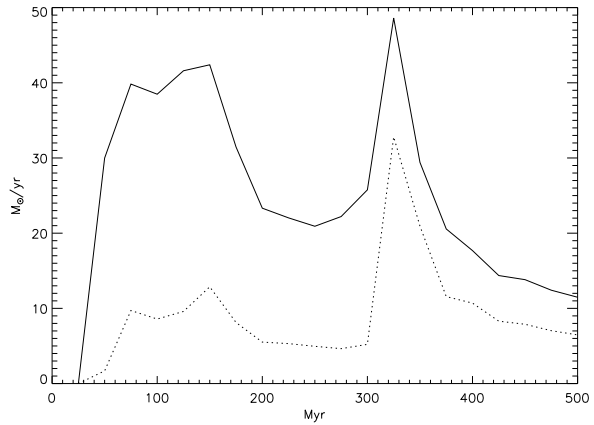,height=6.cm,angle=0}
\caption{The global star formation rate in the merging system (solid line) compared to the star formation
rate in the nucleus of galaxy 1 (dotted line).}
\end{figure}
\begin{figure}
\psfig{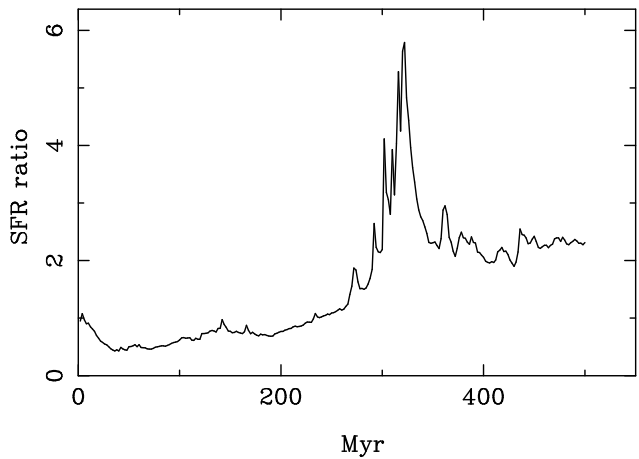}
\caption{Ratio between star formation rate in the merger simulation and star formation rate for an isolated
galaxy as a function of time.}
\end{figure}

\subsection{The dusty torus}
The second important finding is the formation of a nuclear starburst ring or dusty torus at the 
galactic centre. 
The large number of particles in the central 300\,pc combined with the use of a tree code enable us to obtain a
higher resolution in the central region.
We have estimated that, even under the most disfavourable assumptions, our resolution limit in the central region is smaller than
40\,pc. Therefore, we are confident to resolve the hydrodynamics of the central 300\,pc region, from the mean properties of
which we calculate the black hole accretion rate semi-analytically (Eq.~3).
The torus in Fig.~5 has an inner radius of $\sim 40\,$pc and an outer 
radius $\sim 250\,$pc with most of the mass at $\sim 130\,$pc. 
The mass-averaged rotation speed and the one-dimensional velocity dispersion within a distance $r_{\rm max}=250\,$pc from the
centre of the remnant are $v_{\rm rot}=140{\rm\,km\,s}^{-1}$ and $\sigma=63{\rm\,km\,s}^{-1}$, respectively.
We have verified that 
$v_{\rm rot}^2\simeq {\rm G}M(r_{\rm max})/r_{\rm max}$, where $M(r_{\rm max})$ is the mass at $r<r_{\rm max}$ and
G is the gravitational constant, which means that the torus is supported by rotation.
The ratio between the thickness and the diameter of the torus is consistent with the $\sigma/v_{\rm rot}$ ratio.
The formation of a central starburst ring such as the one described in this article
has already been reported by \citet{heller_shlosman94}. 
They simulated the dynamical evolution of globally unstable galactic discs and
discussed how supernova explosions, radiation driven winds from massive stars and the presence of supermassive black holes
affect the dynamical evolution of the starburst ring.
Our simulation does not model these effects, which may play an important role in the evolution of the central region. 
We observe the formation of a well defined torus after the nuclei of the two galaxies have merged, but
after $\sim 50\,$Myr the torus is no longer recognisable in an obvious manner.
At that point the structure of the galactic nucleus changes from a torus into a mini-spiral with
asymmetric spiral arms as more gas and stars fall into the central region.
Less obvious tori are also seen before the two nuclei merge.

\begin{table}
\caption{Virtual observations of the AGN at different viewing angle}
\begin{tabular}{c|c|c|c}
$\theta$ (degree) & $<N_{\rm H}> (10^{24}{\rm\,cm}^{-2})$  & $Z/Z_\odot$ & $M_B({\rm AGN})$ \\
\hline
       0&     0.196349&      1.89471&     -23.7\\
       3&     0.294102&      1.49658&     -23.7\\
       6&     0.292391&      1.81359&     -23.7\\
       9&     0.287282&      1.71863&      376\\
      12&     0.216616&      1.89728&     -23.7\\
      15&     0.155719&      2.10944&     -23.7\\
      18&     0.166973&      1.74387&      376\\
      21&     0.478099&      1.32532&     -23.7\\
      24&     0.339831&      1.37435&     -23.7\\
      27&     0.340767&      1.75315&      376\\
      30&     0.329107&      1.59405&      376\\
      33&     0.361234&      1.54440&      376\\
      36&     0.599168&      1.60186&      376\\
      39&     0.359224&      1.65867&     -23.7\\
      42&      1.14970&      1.64893&      735\\
      45&     0.351887&      1.81494&      376\\
      48&     0.769909&      1.77040&      521\\
      51&    0.0939574&      1.72862&     -23.7\\
      54&     0.666031&      1.92951&      490\\
      57&     0.581031&      1.75790&      385\\
      60&     0.505869&      1.63132&      376\\
      63&     0.693351&      1.67498&      441\\
      66&     0.436236&      1.64965&     -23.7\\
      69&     0.640693&      1.72126&      417\\
      72&      1.59603&      1.86413&      1166\\
      75&      1.06212&      2.01813&      834\\
      78&     0.879192&      1.78262&      603\\
      81&     0.997245&      2.03174&      787\\
      84&      1.00148&      1.66960&      645\\
      87&      1.05674&      1.80867&      741\\
      90&      1.45602&      1.81765&      1035\\
\hline
\end{tabular}
\end{table}

The nuclear region is very dusty. We have assumed that the torus is made of clouds with 
$n_{\rm H}^{\rm cl}=10^{24}{\rm\,cm}^{-2}$ (like the Orion nebula) and we have conducted 
virtual observations of the AGN $350\,$Myr after the simulation began.
We have observed the AGN at inclination angles from pole-on ($\theta=0^\circ$) to edge-on 
($\theta=90^\circ$). The azimuthal angle $\phi$ was selected randomly at each observation.
In Table 2 we give the results  of this Monte Carlo experiment showing the smoothed column
density $<n_{\rm H}>$ and the extinguished AGN blue magnitude $M_B$ for each observation.
$M_B=-23.7$ is the absolute blue magnitude of the unobscured AGN and $M_B=376$ corresponds to 
one cloud on the line of sight. 
 
The average column density of neutral hydrogen on the line of sight to the
AGN is fairly high at all viewing angles and grows by an order of magnitude
when we pass from a pole-on to an edge-on view. 
If the dusty torus is composed of clouds with high optical depth, then 
the AGN appears as a Seyfert 1 when it is observed pole-on 
and as a Seyfert 2 when it is observed edge-on. Table 2 shows that
there are a few exceptions to this rule due to presence of clouds in the 
polar regions and holes in the dusty torus.
Table 2 also shows that the metallicity of the dusty torus is about twice the
solar value due to intense star formation activity in the torus.

\subsection{Quasar hosts}

Quasar host galaxies have been investigated according to both morphological and spectral
criteria. 
\begin{figure}
\psfig{figure=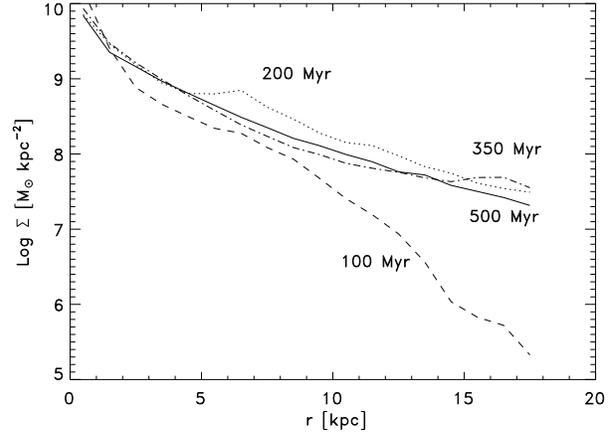,height=6.cm,angle=0}
\caption{Surface density profile of the 1st galaxy 100 (dashed line), 200 (dotted line), 
350 (dashed-dotted line) and 500 (solid line) Myr after the simulation has started.}
\end{figure}
\begin{figure}
\psfig{figure=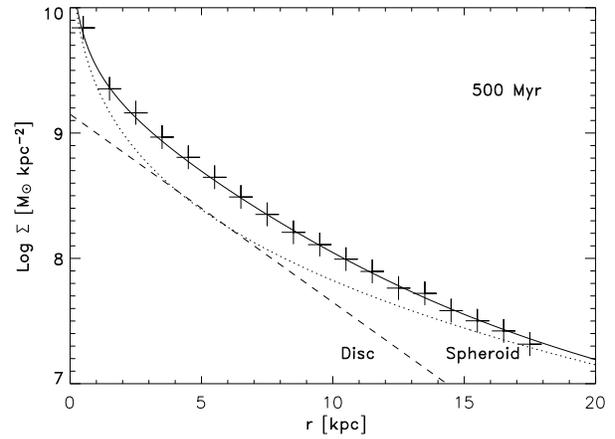,height=6.cm,angle=0}
\caption{Surface density profile of the merger remnant 500 Myr after the simulation has started
(crosses). The solid line fit is the sum of 
a de Vaucouleurs profile $\propto {\rm exp}[-(r/r_{\rm e})^{1/4}]$ (dotted line) and
an exponential disc $\propto {\rm exp}(-r/r_{\rm d})$ (dashed line).
The fit shown on this diagram corresponds to
$M_{\rm spheroid}=1.34\times 10^{11}M_\odot$, $M_{\rm disc}=7.4\times 10^{10}M_\odot$.}
\end{figure}
\begin{figure}
\begin{center}
\centerline{\psfig{figure=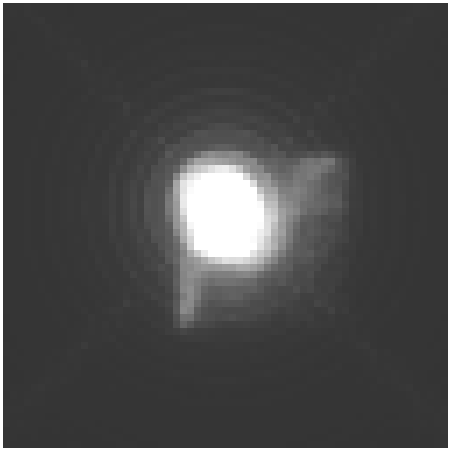,height=6.cm,angle=0}}
\end{center}
\caption{Mock Hubble Space Telescope image of the system 500$\,$Myr 
after the simulation has started. 
The system is assumed to be at $z=0.27$
and is observed at a viewing angle from which it is possible to see the central engine,
hence the diffraction rings.
The imaging model mimics the 814$\,\mu$m filter on the Wide Field Camera. }
\end{figure}
We have analysed the morphology of the merger remnant by using IRAF to fit elliptical 
isophotes to  R-band mock images.
We compared the photometric profiles extracted 
by this procedure with de Vaucouleurs and exponential disc profiles.
We repeated the same analysis by using mass surface density profiles determined by
computing the surface density in circular coronae around the galactic centre.
Both methods gave similar results in terms of the bulge/disc decomposition.
Fig.~8 shows how the surface density profile evolved from $t=100\,$Myr to $t=500\,$Myr.
We see a transition from a more exponential disc-type to a more de Vaucouleurs-type 
profile. In Fig.~9 we show a spheroid plus disc decomposition of the surface density
profile at 500\,Myr. This decomposition corresponds to 
$M_{\rm spheroid}=1.34\times 10^{11}M_\odot$ and $M_{\rm disc}=7.4\times 10^{10}M_\odot$.
Fig.~10 shows how the merger remnant would appear in a  Hubble Space Telescope image
if it was at a redshift of 0.27 and was photographed with the
Wide Field Camera mounted with the $814{\rm\,\mu m}$ filter.
The AGN is visible and has a luminosity of $M_B=-22$.
We clearly see two tails of debris falling on the merger remnant from opposite directions.

Fig.~11 shows how the optical and infrared spectrum evolves during the merger. 
The solid lines show the total spectrum while the dashed lines show the spectrum without the AGN.
The optical spectrum is produced by direct light from stars and from the AGN.
The infrared/sub-mm spectrum is produced by thermal dust heated by stars and by the AGN.
The AGN contributes to the infrared and sub-millimetre emission
independently of the viewing angle because most of the AGN power is absorbed and thermalized
whether the clouds are on the line of
sight or not (e.g. \citealp{sazonov_etal04}).
On the other hand, the direct optical emission from the AGN is
totally extinguished in most cases.


In Fig.~11, B-band corresponds to ${\rm Log}(\lambda/\mu{\rm m})\simeq -0.37$.
At all times, the old stellar populations 
the most important contribution to the optical spectrum comes from the old stellar population
preexisting the merger,
not only because most of the stellar mass is in old
stars, but also because young stars are heavily obscured by dust.
However, as the merger progresses, clear signals of star forming activity emerge
from the infrared and ultraviolet parts of the spectrum. 
The presence of an AGN raises the infrared flux substantially by
increasing the temperature of the dust in the central starburst ring
from  $43\,$K to $63\,$K.

\section{Discussion and conclusion}

We have developed a tool that can simulate real galaxy images to a very high degree of
realism.
Our study confirms the powerful effect of mergers on galactic morphologies 
and the accumulation of cold gas in the galactic centre that were seen in previous studies
(i.e. \citealp{barnes_hernquist96,mihos_hernquist96}).
The gas that concentrates in the galactic centre forms a starburst ring, which can fuel an AGN.
\citet{heller_shlosman94} have already encountered this configuration in SPH simulations
of globally unstable isolated disc galaxies.
They have made simulations with and without supernova feedback.
Without feedback, gas has a much stronger tendency to fragment into few large clumps, but the effect
on the final black hole mass is not significant.
\begin{figure*}
\noindent
\begin{minipage}{8.6cm}
  \centerline{\hbox{
      \psfig{figure=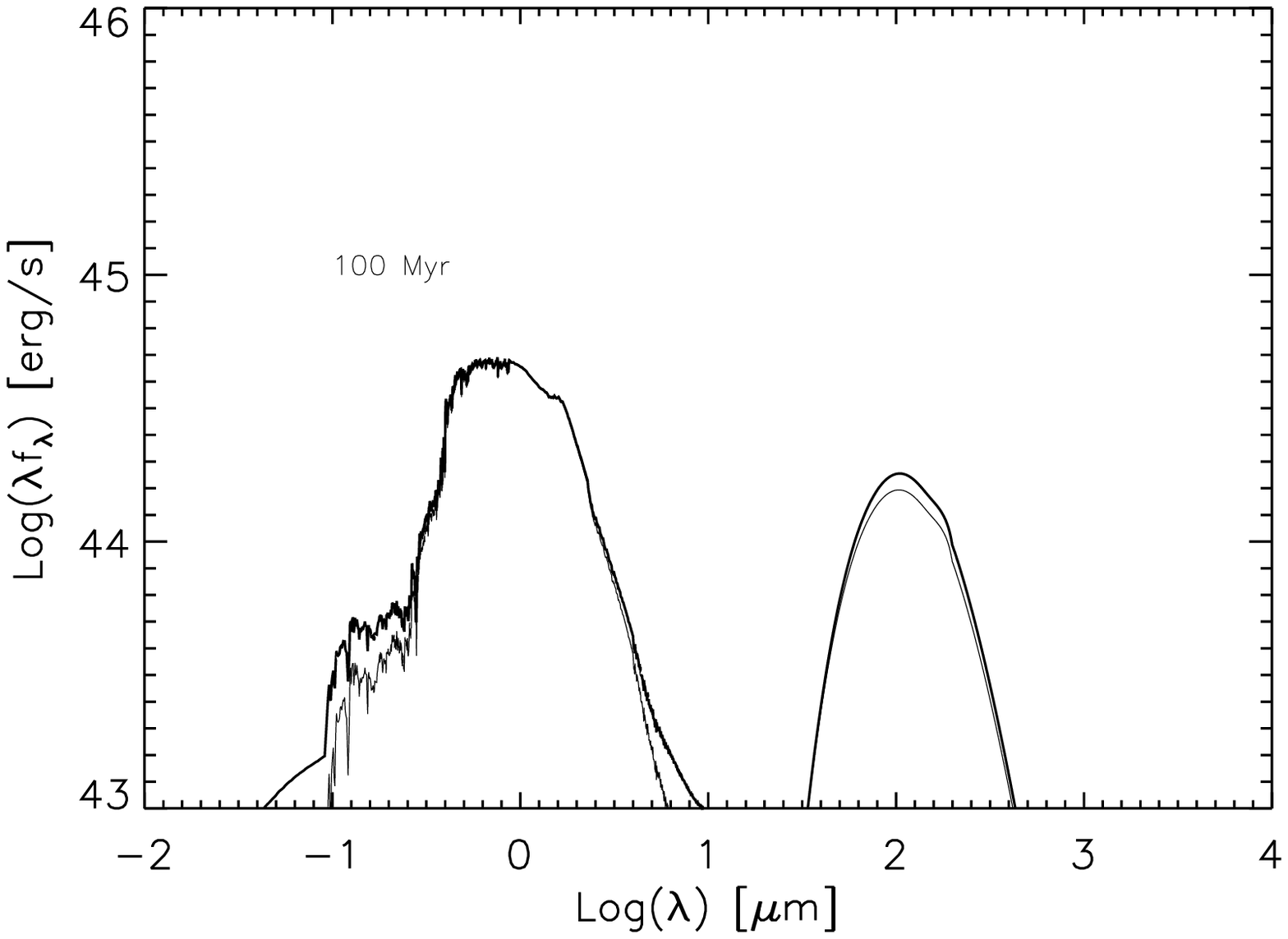,height=6.5cm,angle=0}
  }}
\end{minipage}\    \
\begin{minipage}{8.6cm}
  \centerline{\hbox{
      \psfig{figure=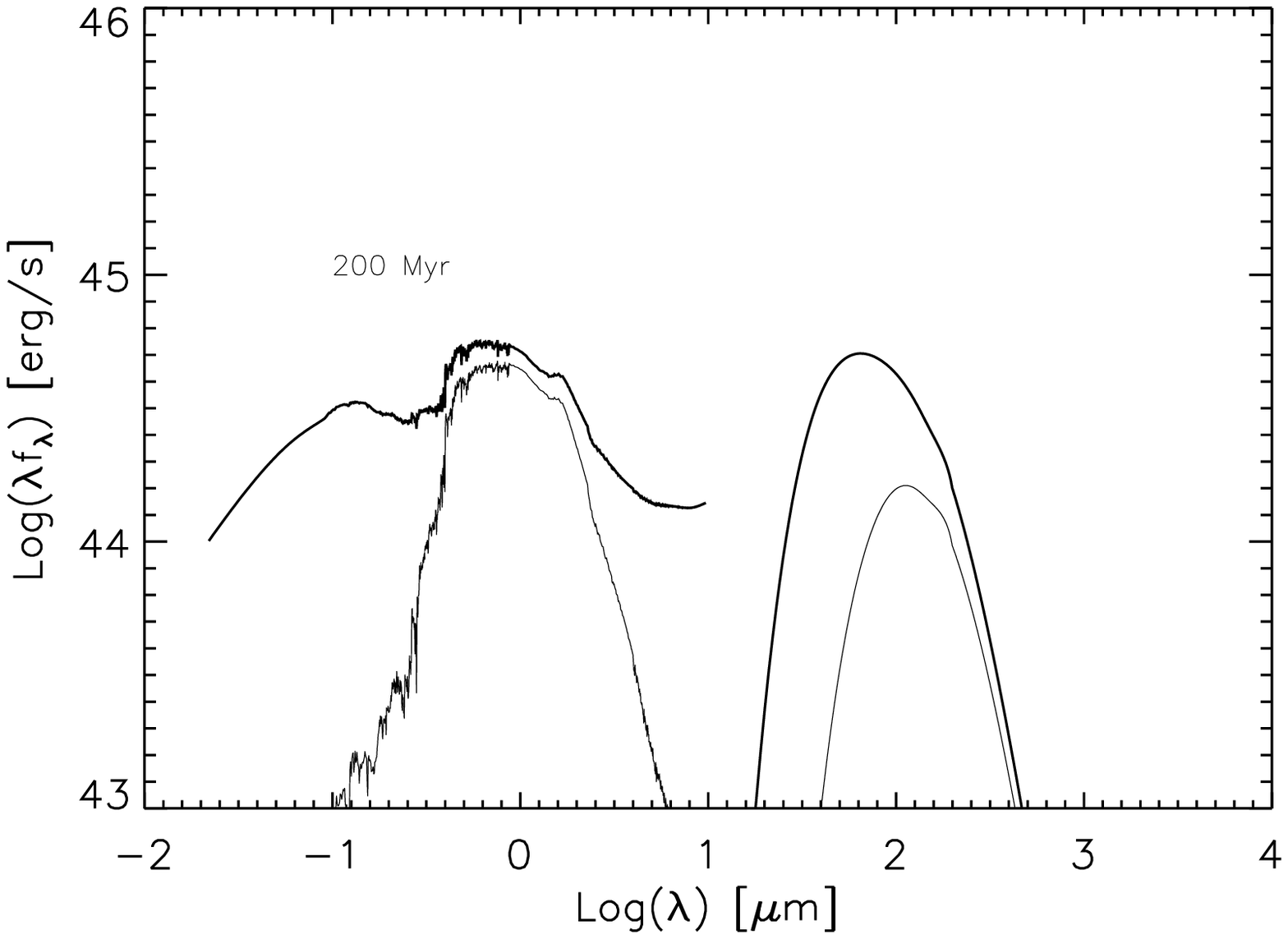,height=6.5cm,angle=0}
  }}
\end{minipage}\    \
\begin{minipage}{8.6cm}
  \centerline{\hbox{
      \psfig{figure=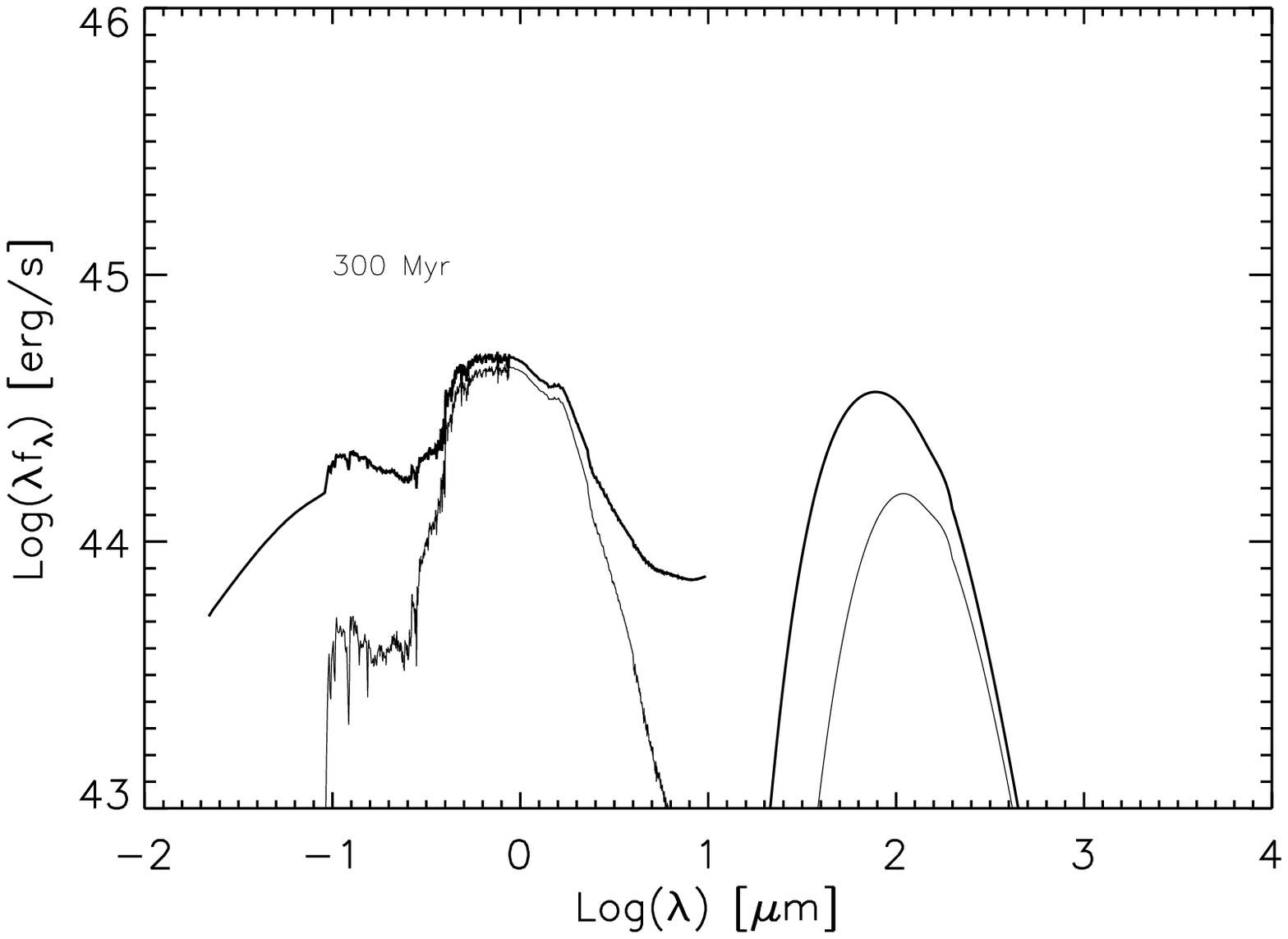,height=6.5cm,angle=0}
  }}
\end{minipage}\    \
\begin{minipage}{8.6cm}
  \centerline{\hbox{
      \psfig{figure=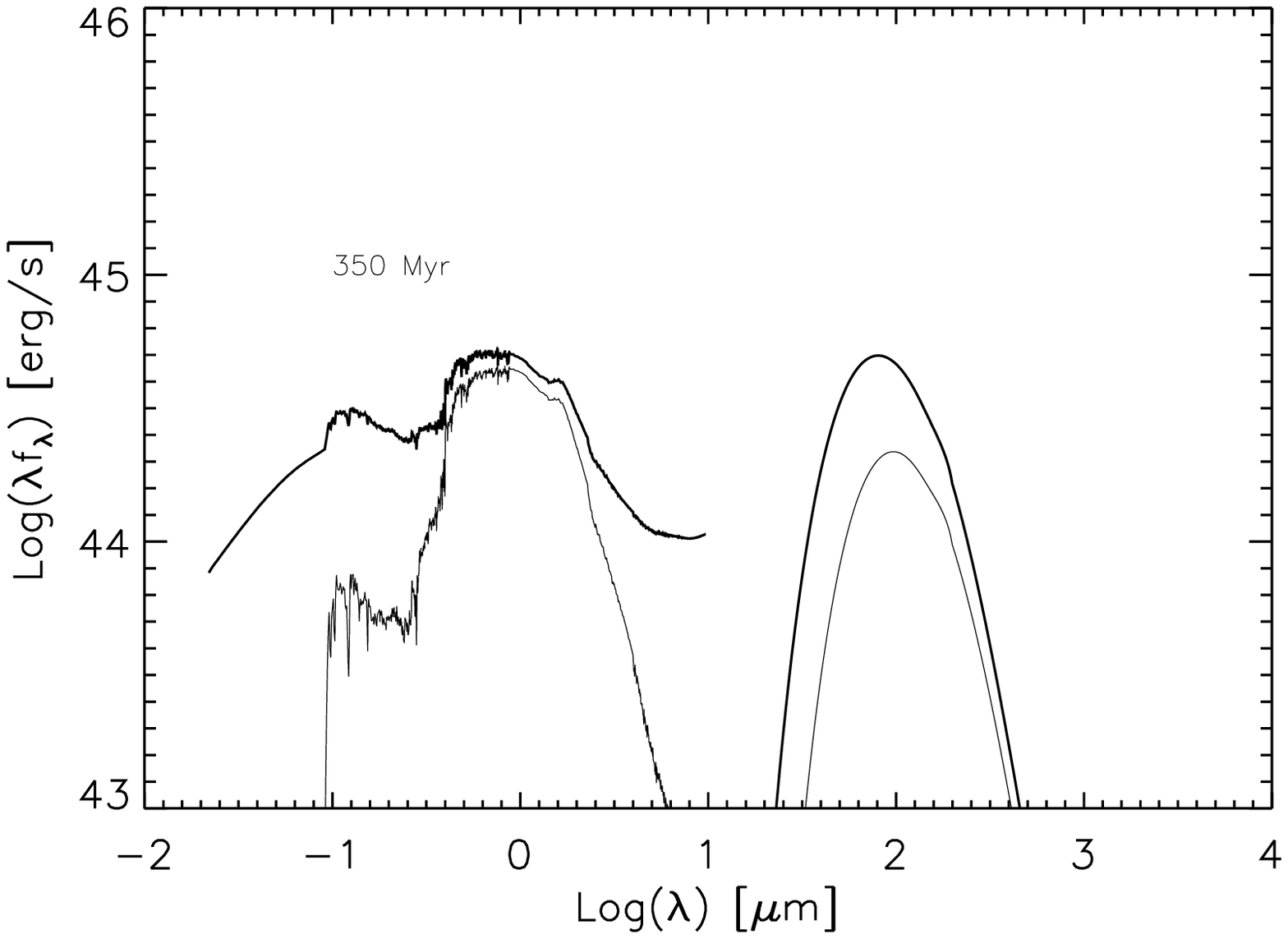,height=6.5cm,angle=0}
  }}
\end{minipage}
\caption{Global spectral energy distribution of the interacting system
100, 200, 300 and 350\,Myr after the simulation has started. 
At optical wavelengths, the thick line (above) corresponds to the case in which in the AGN 
is visible and the thin line (below) to the case in which it is obscured. 
At infrared wavelengths,
the thick line (above) corresponds to dust heated by stars and by the AGN. The thin line
(below) corresponds to dust heated by stars only.} 
\end{figure*}
We are aware that we lack the resolution to investigate the distribution of gas in the starburst ring and
the black hole accretion rate from the SPH simulation, although our resolution length is at least three times
smaller than the characteristic radius of the torus.
Instead, we study these aspects by applying a semi-analytic treatment to the outputs of the simulation.

Even under conservative assumptions for the fuelling rate, the accretion of gas increases
the black hole mass by a factor of $\gsim 3$.
Another factor of 2 is gained by merging the two black holes, so that the
black hole starts with a mass of $3\times 10^7\,M_\odot$ and reaches a mass of
$\sim 1.8\times 10^7\,M_\odot$ by the end of the simulation.
We have simulated a relatively gas-poor merger.
The starburst would be much stronger if the discs had a higher gas fraction.
So would the growth of the central black hole. 

This increase in black hole mass is comparable to the increase in bulge mass from
$2.5\times 10^{10}M_\odot$ to $1.34\times 10^{11}M_\odot$, corresponding to a factor of
$\simeq 5.4$. Therefore, 
with our normalisation of the black hole accretion rate in Eq.~(3), the black hole mass to
bulge mass ratio is comparable at the beginning and at the end. 
The observations suggest that the scatter in the black hole mass to bulge mass relation is 
small (see the Introduction and also \citealp{haering_rix04}).
That means that our normalisation of the black hole accretion rate 
is not far away from the appropriate value because it
preserves the $M_\bullet/M_{\rm spheroid}$ ratio.

The star formation rate in the nuclear region closely follows the global star formation
of the host galaxy. It has two peaks, which coincide with the first and the 
second pericentre passage. 
Fig.~7 compares the star formation in our merger simulation with the star formation rate
in a simulation in which there is only one isolated galaxy. Manifestly, the dynamical 
instability deriving from the initial conditions for the discs, which are thin and 
axisymmetric, exaggerate the pace at which spiral arms form and gas fragments into knots.
Nevertheless, we believe that the filamentary lumpiness of star formation 
evident in Fig.~5 reflects a real astrophysical phenomenon.
Therefore, the first peak in Fig.~6 is artificial.
Instead, the second peak is physical
and corresponds to the actual merger. The starburst triggered by the merger is $\sim 6$ times
more intense than quiescent star formation. Its duration is of the order of 10$\,$Myr.
This peak coincides with maximum AGN activity and occurs when the merger remnant begins 
to look like a lenticular galaxy both colour-wise and morphology-wise (Fig.~4).

Young stars are concentrated in the nuclear starburst ring. Therefore, it is difficult to
disentangle their light from the AGN point-spread-function (point sources appear to have spikes
connected by a spider web owing to instrumental diffraction, Fig.~10). However, this is a merger with a low
gas fraction.
It is possible that, in the case of a gas-rich merger, star formation is no longer confined to the nuclear
region, but distributed throughout the host galaxy \citep{kauffmann_etal03}.

We observe the formation of a dusty torus, which obscure the AGN from most 
viewing angles. The torus has as an opening angle of $30^\circ-40^\circ$ and its properties are
consistent with expectations of unified model (see \citealp{rowan-robinson95}) if the
torus is made of clouds with $n_{\rm H}\sim 10^{24}{\rm\,cm}^{-2}$ (like the Orion nebula).
Our model predicts that in some case the AGN is obscured even from a face-on viewing angle 
owing to the presence of clouds in the polar regions, while,
in other cases, we can still see the AGN from nearly equatorial viewing-angles
because there are holes in the dust distribution.

\section*{Acknowledgements}

A. Cattaneo acknowledges the support of the European Commission through a Marie Curie Research Fellowship.
He also wishes to thank B. Guiderdoni and M. Rowan-Robinson for useful conversation,
F. Durret for help with the IRAF software and 
G. Mamon for comments on the manuscript.




\bibliographystyle{mn2e}
\bibliography{references}
\end{document}